\documentclass[twocolumn,12pt,tighten]{aastex631}
\usepackage{amsmath,amssymb}

\usepackage{lipsum}
\usepackage{graphicx}
\usepackage{subfigure}
\renewcommand{\l}{\left}
\renewcommand{\r}{\right}

\hypersetup{linkcolor=blue,citecolor=blue,filecolor=cyan,urlcolor=magenta}
\usepackage{hyperref}
\usepackage{cleveref}
\usepackage{comment}

\usepackage[normalem]{ulem}

\begin{document}
\newcommand{\onePETattwon}{$27^{+7}_{-4}$}
\newcommand{\onePETatfiven}{$445^{+89}_{-81}$}
\newcommand{\onePAplusatfiven}{$465^{+118}_{-111}$}
\newcommand{\onePAplusatftwon}{$29^{+11}_{-6}$}
\newcommand{\onecsETattwon}{$0.30^{+0.08}_{-0.05}$}
\newcommand{\onecsETatfiven}{$0.77^{+0.03}_{-0.04}$}
\newcommand{\onecsAplusatfiven}{$0.78^{+0.04}_{-0.07}$}
\newcommand{\onecsAplusatftwon}{$0.32^{+0.14}_{-0.09}$}

\newcommand{\tenPETattwon}{$27^{+2}_{-2}$}
\newcommand{\tenPETatfiven}{$453^{+22}_{-23}$}
\newcommand{\tenPAplusatfiven}{$468^{+42}_{-42}$}
\newcommand{\tenPAplusatftwon}{$28^{+4}_{-3}$}
\newcommand{\tencsETattwon}{$0.31^{+0.02}_{-0.01}$}
\newcommand{\tencsETatfiven}{$0.78^{+0.01}_{-0.00}$}
\newcommand{\tencsAplusatfiven}{$0.78^{+0.01}_{-0.02}$}
\newcommand{\tencsAplusatftwon}{$0.33^{+0.05}_{-0.04}$}

\newcommand{\allPETattwon}{$25^{+3}_{-3}$}
\newcommand{\allPETatfiven}{$424^{+47}_{-43}$}
\newcommand{\allcsETattwon}{$0.28^{+0.04}_{-0.02}$}
\newcommand{\allcsETatfiven}{$0.76^{+0.02}_{-0.02}$}

\newcommand{\onePETastroattwon}{$25^{+4}_{-3}$}
\newcommand{\onePETastroatfiven}{$419^{+58}_{-57}$}
\newcommand{\onePAplusastroatfiven}{$404^{+72}_{-69}$}
\newcommand{\onePAplusastroattwon}{$24^{+5}_{-4}$}
\newcommand{\onecsETastroattwon}{$0.28^{+0.05}_{-0.03}$}
\newcommand{\onecsETastroatfiven}{$0.76^{+0.02}_{-0.03}$}
\newcommand{\onecsAplusastroatfiven}{$0.75^{+0.03}_{-0.05}$}
\newcommand{\onecsAplusastroattwon}{$0.26^{+0.07}_{-0.05}$}

\title{{Constraining nuclear parameters using Gravitational waves\\ from 
f-mode Oscillations in Neutron Stars }}

\author[0000-0002-2526-1421]{Bikram Keshari Pradhan}
\email{bikramp@iucaa.in}
\author[0000-0001-8129-0473]{Dhruv Pathak}
\author[0000-0002-0995-2329]{Debarati Chatterjee}
\affiliation{Inter University Centre for Astronomy and Astrophysics, Pune, Maharastra, 411007,India}

\begin{abstract}

 Gravitational waves (GW) emanating from unstable quasi-normal modes in Neutron Stars (NS) could be accessible with the improved sensitivity of the current GW detectors or with the next-generation GW detectors and, therefore, can be employed to study the NS interior. Assuming f-mode excitation in isolated pulsars with typical energy of pulsar glitches and considering potential f-mode GW candidates for A+ (upgraded LIGO detectors operating at 5th observation run design sensitivity) and Einstein Telescope (ET), we demonstrate the inverse problem of NS asteroseismology within a Bayesian formalism to constrain the nuclear parameters and NS Equation of State (EOS). We describe the NS interior within relativistic mean field formalism. Taking the example of glitching pulsars, we find that for a single event in A+ and ET, among the nuclear parameters, the nucleon effective mass ($m^*$) within 90\% credible interval (CI) can be restricted within $10\%$ and $5\%$, respectively. At the same time, the incompressibility ($K$) and the slope of the symmetry energy ($L$) are only loosely constrained. Considering multiple (10) events in A+ and ET, all the nuclear parameters are well constrained, especially $m^*$, which can be constrained to 3\% and 2\% in A+ and ET, respectively. Uncertainty in the observables of a $1.4M_{\odot}$ NS such as radius ($R_{1.4M_{\odot}}$), f-mode frequency ($f_{1.4M_{\odot}}$), damping time ($\tau_{1.4M_{\odot}}$) and a few EOS properties including squared speed of sound ($c_s^2$) are also estimated.
\end{abstract}


\section{Introduction}\label{sec:introduction}

One of the main areas of interest in the current research is the enigmatic interior of neutron stars (NS). NSs provide a natural laboratory to study nuclear matter under extreme conditions such as high magnetic field and rotation currently unreachable by terrestrial experiments~\citep{Lattimer2021}. For cold NSs, the interior is described by the pressure density relationship: known as equation of state (EOS). The NS EOS plays a crucial role in connecting the interior composition with the NS structure properties measured using electromagnetic observations. Furthermore,  the detection of gravitational waves from the binary neutron star event GW170817 by the LIGO and Virgo collaboration ~\citep{AbbottPRL119,AbbottPRL121,Abbottprx}  opened up a new window to multimessenger astronomy ~\citep{AbbottAJL848}. In recent works,  different astrophysical and other nuclear observations have been considered jointly to describe NS physics ~\citep{Most2018,Biswas2021,Bauswein_2017,Essick2021,Ghosh2022,Huth2022}. With the increasing sensitivity of the GW detectors or with the next generation GW detectors such as Einstein Telescope (ET) \citep{Hild_2011}, Cosmic Explorer (CE) ~\citep{CE}, and the Neutron star Extreme Matter Observatory (NEMO)~\citep{nemo}, the NS stellar quasinormal modes (QNM) can be detectable which will provide an excellent opportunity to understand the NS interior.

NS can have different fluid QNMs classified according to the restoring force, such as fundamental f-mode, pressure p-mode (pressure is the restoring force), and g-mode (buoyancy is the restoring force). 
 The QNMs can be excited in different physical scenarios in isolated as well as in binary systems ~\citep{Keer2014,Abbott_2019,Ferrari2003,Andersson2018,Steinhoff2016,Stergioulas2011,Takami2015,Ho2020}. Among the QNMs of NS, f-mode strongly couples with the {GW emission}  and also the mode frequency falls under the detectable frequency range of the current and next-generation GW detectors. Furthermore, different works have shown that the g-modes are less significant than f-modes for GW emission ~\cite {Ferrari2003,Lai1999,kruger2015}, leading us to focus on the f-mode asteroseismology. 

The idea of NS asteroseismology involving NS QNMs started with the pioneering work of ~\citet{Andersson96,Andersson98}. The key idea is to infer the NS global properties such as mass ($M$), radius ($R$), the moment of inertia ($I$), rotational frequency ($\nu$) etc., from the observation of mode characteristics (mainly frequency and damping time) by use of EOS-independent relations or Universal relations (UR) proposed among the mode parameters and NS observables. The URs were further improved in ~\citet{Tsui2005,Lioutas2017,Sotani_Bharat2021}. Though the URs are proposed to be EOS-independent, some URs show EOS-dependence. The URs have been studied in subsequent works with different assumed NS EOS models or different NS compositions ~\citep{Benhar,Sotani2011,Sandoval,Flores2017,Jaiswal,Pradhan2021,das2021,Kumar_2023,Zhao2022,Aguirre2022,Sandoval2022,Pradhan2022}. Though initial URs involved the compactness ($M/R$) and mean density of the star, later works have shown that there exist universality of mode parameters with the inertia parameter ~\citep{Lau_2010,Chirenti2015}, the tidal deformability parameters ~\citep{Chen2014,Pradhan2023}, and with the nuclear parameters ~\citep{Sotani2021,Kunjipurayil2022}, which are helpful in different scenarios in NS asteroseismology problem.  Additionally, a significant effort has been put forward to investigate the impact of rotation on the QNMs and the NS asteroseismology or URs  ~\citep{Zink2010,Kastaun2010,Gaertig2011a,Yoshida2012,Passamonti,Doneva,Pnigouras2016,Yazadjiev2017,Rosofsky2019,Kruger2020b,Kruger2020a}.

 The QNM frequency and damping time contain the information on NS composition; hence the detection of NS QNMs can be used inversely to constrain the NS interior: the inverse asteroseismology. The uncertainty in mode parameters that might result from a detection has been discussed in ~\citet{Kokkotas2001}. The inverse problem of NS asteroseismology to reconstruct the NS observables for rotating NS has been analyzed in ~\citet{Volkel2021}, where two different methods have been used; one involves using URs and another based on the assumption of a true EOS. Additionally, in ~\citet{Volkel2022}, the authors have investigated the inverse problem involving f-modes of rotating NSs and other observations to constrain the nuclear EOS by considering a polytropic description of NS matter. In ~\citet{Zhang_2021}, the authors try to constrain the nuclear symmetry energy assuming a synthetic error on the f-mode frequency of a NS of known mass. However, the inverse problem may not always require prior knowledge of the NS mass. In the mentioned works, an assumed error is considered for mode frequency, and the uncertainty of the damping time was ignored.

Sudden increase in the spin frequency of the pulsars is known as a glitch whereas its sudden decrease is referred to as anti-glitch. A significant number of explorations have been made regarding the relationship between glitches and f-mode excitations in recent works~\citep{Abbott_2019,AbbottPRD104}. There have also been recent works on f-mode based GW searches from the LVK collaboration. Apart from the above-mentioned \citet{AbbottPRD104} which is regarding the all-sky search for short GW bursts, \citet{AbbottLVK2022} search for the short-duration GW signals emitted by f-modes from magnetar bursts, and \citet{AbbottFRB2022} search for GW signals associated with fast radio bursts (FRBs), where the injected waveforms they use are chosen to model the NS f-modes. Though there is no confident detection  of GW events corresponding to NS f-mode as reported by the mentioned searches, they have improved the upper  strain limit for the potential sources resulting from NS f-modes.
\citet{Lopez2022} consider the GW generating f-modes excited by NS glitches and update the all-sky upper limits from the LIGO-VIRGO O3 run for those GW signals from f-mode events and { also concluded that the sources can be localized precisely even without electromagnetic counterpart in the future GW observations}.  \citet{Yim2023} recently proposed a model for describing observed small spin-ups and spin-downs in the corresponding glitch and anti-glitch candidates. In this model, the changes in the spin frequency are attributed to the excitation and decay of non-axisymmetric f-modes, with mode excitation amplitude being the only free parameter. In a recent work, \citet{Moragues2022} estimate the detection prospects of long-duration quasi-monochromatic transient GW signals from glitching pulsars.

{Different phenomena, such as newborn NSs~\citep{Ferrari2003}, star quakes ~\citep{Kokkotas2001,Mock_1998}, magnetars~\citep{Abbott_2019,AbbottLVK2022}, and the pre-merger ~\citep{Steinhoff2016,Andersson2018} and post-merger stages of a NS in binary~\citep{Shibata1994,Stergioulas2011,Bauswein2012}, can result in an f-mode excitation. In order to perform estimates of the detectability, in this work we consider GW signals resulting from the excitation of f-modes in isolated pulsars, where the mode is excited to a level corresponding to the energy associated with typical pulsar glitches.  The motivation for the glitch model comes from present f-mode GW searches by LIGO-VIRGO observations. While discussing NS seismology~\citep{Kokkotas2001,Andersson2001,Andersson2021} and the detectability~\citep{Ho2020,Abbott_2019,AbbottPRD104,AbbottFRB2022,AbbottLVK2022} of transient f-mode GW signal, the assumption of f-mode excitation with an energy similar to that of typical pulsar glitches has been widely considered, even though a clear explanation for the connection between the pulsar glitches and the mode excitation is yet to be found. By taking into account typical glitch energy and focusing on glitching pulsars, the problem becomes calculable and straightforward. Additionally, for isolated NSs, the population and other characteristics (such as sky position) are well known, and future observations of the current or future f-mode GW searches may involve the glitch model in a manner similar to that of current searches  as carried out for the earlier observations.}

 
 In this work, we follow the methodology of ~\citet{Ho2020} and under the assumption that the f-modes are excited in glitching pulsars with typical pulsar glitch energy, we estimate the possible detectable candidates for the following GW observation runs or for next-generation GW detectors. For the potential sources, we estimate the uncertainty on the measurements of f-mode frequency ($f$) and damping time ($\tau$) and use the ($f,\tau$) posterior in a Bayesian formalism to constrain the nuclear parameters or the  NS EOS. In Section \ref{sec:methodology}, we discuss the methodology of this work, including the GW modeling, estimation of possible candidates, the EOS model, and the Bayesian formalism to constrain the nuclear parameters. We present our results in Section ~\ref{sec:result} and summarise our conclusions in Section ~\ref{sec:conclusion}.

\section{Methodology} \label{sec:methodology}
\subsection{{Model for f-mode GWs}} \label{sec:pulsars}
The GW  resulting  from f-mode oscillations can be modelled as a damped sinusoidal ~\citep{Ho2020,Kokkotas2001}, 
\begin{equation}\label{eqn:gwwaveform}
        h(t)=h_0e^{-(t-t_0)/\tau} \sin{\l[2\pi f (t-t_0)+\phi\r]} ,\ \mathrm{for} \ t\geq t_0
\end{equation}
where, 
\begin{equation}\label{eqn:h0}
h_{0}= 4.85  \times 10^{-17} \sqrt{\frac{E_{\rm gw}}{M_{\odot}c^2}} \sqrt{\frac{0.1 {\rm sec}}{\tau_f}} \frac{1 \rm{kpc}}{d}\left(\frac{1 \rm{kHz}}{f}\right)~.
\end{equation}
  
To estimate observational uncertainty, one needs the value of $E_{\rm gw}$. Different scenarios could exist where the f-mode can be excited in isolated NS, including NS glitches, a sudden phase transition leading to a mini-collapse, or even newborn neutron stars~\citep{Kokkotas2001}. Now following the assumption of ~\citet{Ho2020}, that the f-mode is excited  to a level corresponding to a pulsar glitch, such that the GW energy
$E_{\rm gw}$ is the  energy of the glitch; one can have~\citep{Ho2020},
\begin{equation}\label{eqn:egw}
    E_{\text{gw}}=E_{\text{glitch}}=4\pi^2I\nu^2 (\frac{\Delta \nu}{\nu})~,
\end{equation}
where $I$ and $\nu$ are the moment of inertia and spin frequency, respectively, where $\frac{\Delta \nu}{\nu}$ is the relative change in spin frequency during a glitch of the glitching pulsar. 

 We consider the data of the glitching pulsars from the Jodrell Bank Glitch Catalogue\footnote{\url{https://www.jb.man.ac.uk/pulsar/glitches.html}} \citep{jbglitch11}. The Jodrell Bank Glitch Catalogue lists each detected glitch's relative spin frequency change. We use the ATNF Pulsar Catalogue\footnote{\url{https://www.atnf.csiro.au/people/pulsar/psrcat/}} \citep{atnf05} to supplement the glitch data with each pulsar's spin frequency $\nu$, distance $d$, and sky position. 

 In this work, we consider the EOS model, PCSB0 from \citet{PRADHAN2023122578}, as the theoretical description for the model is based on the RMF model (for which we are aiming to constrain the nuclear parameters), and it also satisfies the state-of-the-art constraints such as Chiral Effective Field Theory and current astrophysical data. 
 Assuming the given EOS model, each glitching pulsar is assigned a random mass from a uniform mass distribution\footnote{One can use a different population model like a uni-modal Gaussian~\citep{Ho2020} or a more realistic bimodal Gaussian distribution from ~\citet{Population}. However, we keep the distribution uniform in the observed and predicted mass range to keep our investigation less biased.} in $[1 M_{\odot}, M_{max}]$, with $M_{max}=2.5M_{\odot}$  as the maximum possible mass of the assumed EOS. For the respectively assigned mass of each pulsar, we assign the non-rotating f-mode frequency ($f$) and damping time ($\tau$) obtained by solving the perturbation equations in the linearized theory of general relativity with the methodology developed in ~\citet{Pradhan2022}. We also assign the corresponding moment of inertia ($I$) for the respective mass resulting from the assumed EOS.  

\begin{figure*}[htbp]
\gridline{\fig{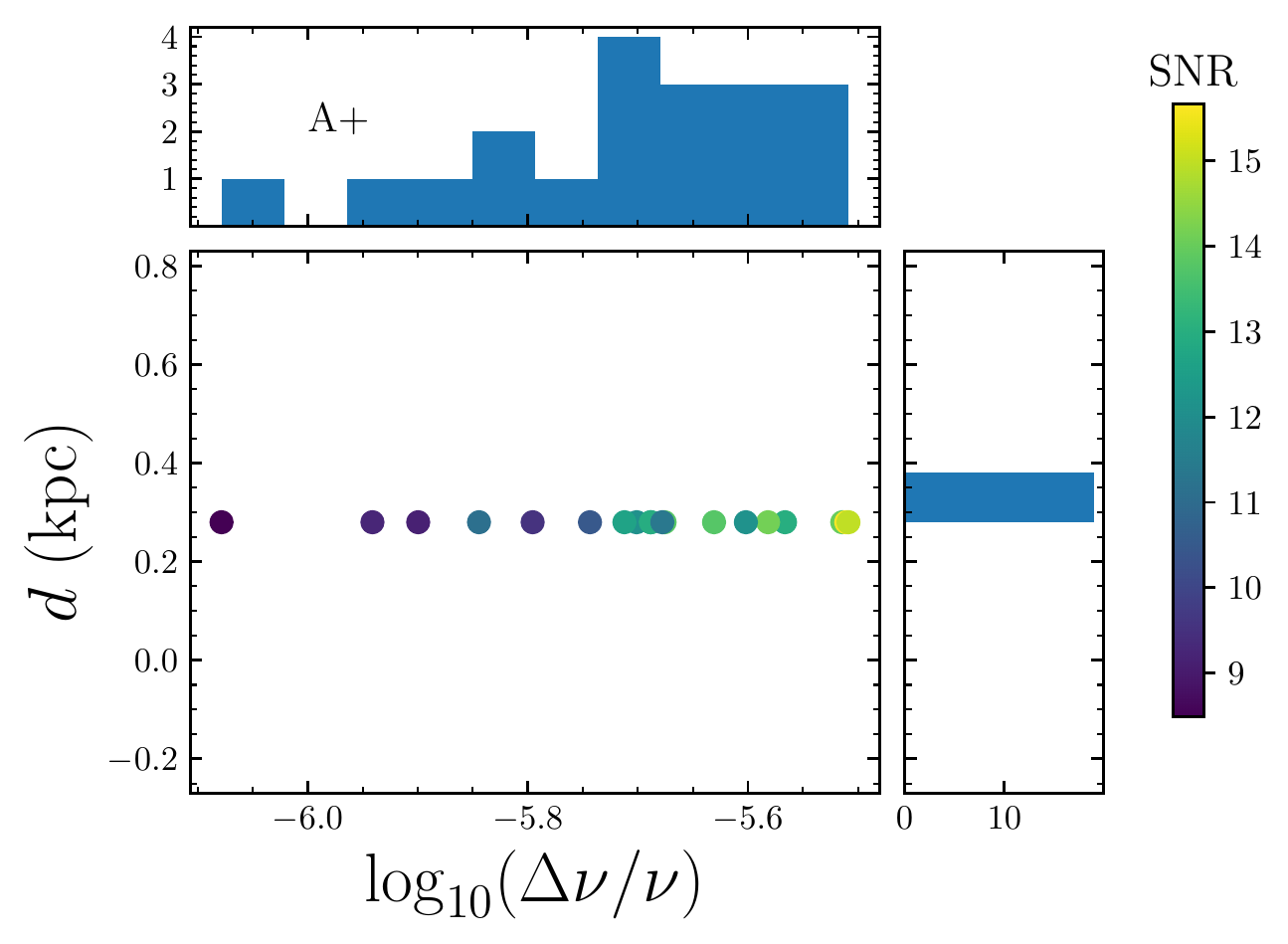}{0.5\linewidth}{(a)}
          \fig{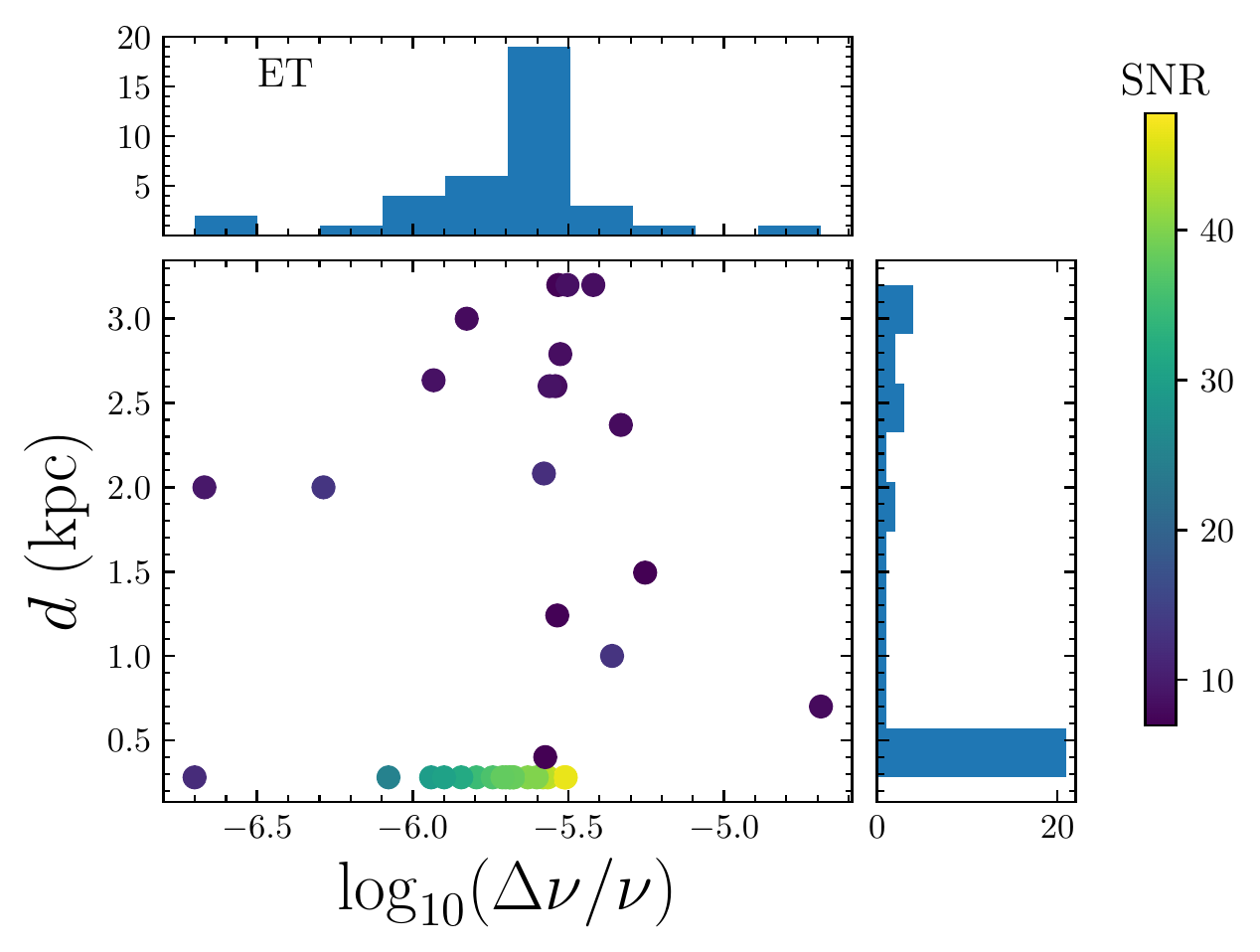}{0.5\linewidth}{(b)}
          }
\caption{Distribution of glitches  in the $d$ and $\frac{\Delta \nu}{\nu}$ plane having  SNR $\geq 8$ in (a) $\rm A+$ configuration and (b) ET configuration. The frequency histogram of the glitches and distances with network SNR $\geq$ 8 are also plotted in the above and right panel respectively. In (a), all glitches shown are from Vela which is also reflected as a single bar in the histogram of (a) corresponding to distance.}
\label{fig:snr}
\end{figure*}

We consider two GW network configurations; in one case, the two LIGO detectors (H1 and L1) are operating with O5 design sensitivity~\cite{Abbott2020}~\footnote{\url{https://dcc.ligo.org/LIGO-T2000012/public}} (referred to as A+). The other case involves the next-generation ET with ET-D sensitivity~\citep{Hild_2011}\footnote{\url{https://dcc.ligo.org/LIGO-T1500293/public}}. We model the GW signal shown in ~\cref{eqn:gwwaveform} using the GW inference package \textbf{bilby}~\citep{bilby} and calculate the signal-to-noise ratio (SNR) for different GW events resulting from f-mode excitation from pulsar glitches. We have considered the detector response resulting from the sky position of the pulsars. In agreement with ~\citet{Ho2020}, we found that in the $\rm A+$ configuration, the  GWs  from  the Vela pulsar  can have SNR $\geq 8$ (due to the proximity of the Vela pulsar). However, in the ET, one can have events  from other pulsars having SNR $\geq 8$. We display the distribution of f-mode GW events  with SNR $\geq 8$ in the  $d$ and $\frac{\Delta \nu}{\nu}$ plane for   $\rm A+$ and ET  configurations in  ~\cref{fig:snr}.

We ignore the rotational correction in this work because recent efforts suggest that the leading order spin correction to the mode frequency is 0.2 ($\nu/\nu_K$ ) ~\citep{Kruger2020a} ($\nu_K$ is the Kepler frequency). Almost all glitching pulsars have a low spin frequency ($\nu <$ 50 Hz). In contrast, the Kepler frequency is $\sim$ 1 kHz, such that they would have f-mode frequency correction $<$ 1\%; this implies that the rotation has a minor effect on a detection event from transient NS f-modes from glitching pulsars. Further, we ignore the fast spinning glitching pulsar (PSR J0613-0200 with $\nu\sim 300$Hz) given its small glitch size and low glitching rate (see Section IV of ~\citet{Ho2020} for detailed discussion).

\subsection{Equation of State: RMF Model}\label{sec:EOS}
The EOS is essentially the relation between the pressure ($p$) and density ($\rho$) (or energy density $\epsilon$) of the NS matter, i.e., $p=p(\epsilon)$. As discussed in ~\citet{OertelRMP}, the two main approaches developed to describe the NS matter involve microscopic  ( or {\it ab-initio} ) description, and the other includes phenomenological (effective theories where parameters are fitted to reproduce the saturation properties of nuclear matter and/or the properties of finite nuclei). The phenomenological models involve  interactions of Skyrme
\citep{Skyrme1956,Skyrme1956b}, Gogny type ~\citep{Decharg1980} and those of the relativistic
approach~\citep{MACHLEIDT19871,Haidenbauer,Shen2010,Cescato1998,Nik2002,Serot1997}. In this work, we employ the Relativistic Mean Field (RMF) model, a relativistic effective field theoretical model successfully applied to a wide range of nuclei, nuclear and NS matter~\citep{Chen2014,Hornick,Ghosh2022,Ghosh2022b,Shirke_2023,PRADHAN2023122578}. 

In the considered  RMF model, the interaction among the NS constituents can be described using a   Lagrangian density where baryons interact through the exchange of mesons. We consider the beta equilibrium  NS  interior composed of nucleons ($n,p$) and leptons ($e,\mu$). The detailed description and methodology of the RMF model used in this work can be found in ~\citet{Hornick}. The RMF model parameters are calibrated to the nuclear parameters at saturation: nuclear saturation density ($n_0$), the binding energy per nucleon  ($E/A$ or $E_{\rm sat}$), the incompressibility  ($K$), the effective nucleon mass ($m^*$), the symmetry energy ($J$) and the slope of symmetry energy ($L$) at saturation density. Hence, for a choice of parameter set $\{\theta\}=\l \{n_0, E_{\rm{ sat}}, K,m^*,J,L\r \}$,  the unique NS EOS is obtained using the methodology described in Section II and III of ~\citet{Hornick}. In the following sections, we assume a particular EOS model ( for injections or assigning the NS properties to the pulsars ) corresponding to the parametrization from ~\citet{Hornick}, which is referred to as PCSB0 EOS model in ~\citet{PRADHAN2023122578} and can be availed from the EOS database \textbf{CompOSE}~\citep{Compose2} \footnote{\url{https://compose.obspm.fr/EOS/292}}. The model parameters for the PCSB0 EOS model are $n_0=0.15 \rm{fm^{-3}}, \  E_{\rm{ sat}}=-16 \rm{MeV},\  K=240 \rm{MeV},\ m^*/m_N=0.65,\ J=32 \rm{MeV}$, and $L=60 \rm{MeV}$.

\subsection{Bayesian Formalism}\label{sec:Bayes}
The process of constraining the nuclear parameters with the preferred EOS model using the f-mode GW asteroseismology involves two steps:
\begin{itemize}
  \item First, for selected candidates, GW signal is injected using ~\cref{eqn:gwwaveform}, and the Bayesian parameter estimation is performed using bilby to obtain the posteriors of signal parameters.   
  \item Further the joint posterior of ($f,\tau$) is used to constrain the nuclear parameters.  
\end{itemize}
  In our analysis through bilby, we consider the detector response due to the sky position and Gaussian noise. This work mostly focuses on the targeted pulsars for which the distance and sky positions are known. Hence, we fix the sky position and distance at their injected value and obtain the posteriors of the mode parameters through the nested sampling algorithm \textbf{dynesty} ~\citep{dynesty} as implemented in bilby. We keep log-uniform prior on $E_{gw}$, a uniform prior on $f\in [800,3000]$ Hz, and a uniform prior in $\tau \in [0.01,0.7]$ s.

For an event `$D$', performing the above with priors  $\pi(E_{gw},f,\tau)$,  will result in a posterior  $P(E_{gw},f,\tau|D)$. However, we use the marginalized posteriors of ($f,\tau$) over other parameters ($E_{gw}$ in this case) for further constraining the EOS parameters. For an event `$D$', the posterior probability for the EOS parameters $\{ \theta\}$ can be obtained using Bayes theorem as,
       \begin{align}\label{eqn:PE_oneevent}
        p(\theta|D)&\propto P(D|\theta)\pi(\theta) & \nonumber \\ 
              &\propto \pi(\theta) \int^{M_{max}(\theta)} dm\   P(D|f(m),\tau(m))\  p(m|\theta)~,  &
     \end{align}
 where $\{\theta\}=\l \{n_0, E_{\rm{ sat}}, K,m^*, J, L\r \}$ are  EOS parameters,  $p(m|\theta)$ is a uniform distribution from a minimum mass $M_{min}=1 M_{\odot}$ to the stable maximum mass $M_{max}$ that can be possible from the EOS model $\{\theta\}$, i.e.,  $p(m|\theta)=U[M_{min}, M_{max}(\theta)]$. In ~\cref{eqn:PE_oneevent}, $f(m)$ and $\tau(m)$ are obtained using the f-Love and $\tau$-Love URs  respectively~\citep{Pradhan2022}. The probability $P(D|f,\tau)$, is constructed from the joint posterior probability $P(f,\tau|D)$ obtained from bilby by dividing the prior $\pi (f,\tau)$, i.e.,
 \begin{align*}\label{eqn:PE_Nevent}
     P(f,\tau|D) &\propto P(D|f,\tau) \pi(f,\tau)~,& \\
     P(D|f,\tau) &\propto \frac{P(f,\tau|D)}{\pi(f,\tau)}~.
 \end{align*}
 For $N$ independent events $\{D\}=\{D_1,D_2,..D_N\}$, the  posterior of the EOS parameters $\{\theta \}$ can be written as,
\begin{align}
    p(\theta|\{D\})&\propto \prod_{i}^{N}P(D_i|\theta)\pi(\theta)~. &
 \end{align}
 We perform the parameter estimation in the Bayesian framework for the EOS parameters as mentioned in  ~\cref{eqn:PE_oneevent,eqn:PE_Nevent} using the python based package \textbf{PyMultinest}\footnote{\url{https://johannesbuchner.github.io/PyMultiNest/}} \citep{Buchner},  based on the nested sampling algorithm . For the analysis presented in Section \ref{sec:result}, a truncated Gaussian prior is considered for each of the nuclear parameters ($x$) with a mean ($\mu_x$) fixed to the parameters of PCSB0 EOS model and the standard deviation ($\sigma_x$)  inspired from \citet{Margueron2018}. We truncate  each normal distribution for  parameter $x$ in the minimum $x_l$ and  maximum $x_u$ range, i.e, the prior of $x$ is taken as ~$\pi(x)=\mathcal{N}(\mu_x,\sigma_x) \ T[x_l,x_u]$. The limits for the nuclear parameters are fixed such that it covers the minimum and maximum value given for the phenomenological EOS models in ~\citet{Margueron2018}. The priors considered are given in the `prior' column of the ~\cref{tab:prior_posterior}. 
 

\section{Results}\label{sec:result}
We discuss the inverse problem to infer the nuclear parameters, first considering a single event in A+ and ET from Vela Pulsar with  the strongest glitch energy  and then considering multiple detectable events in A+ and ET.
\subsection{ Single event }
\begin{figure}
    \centering
    \includegraphics[width=\linewidth]{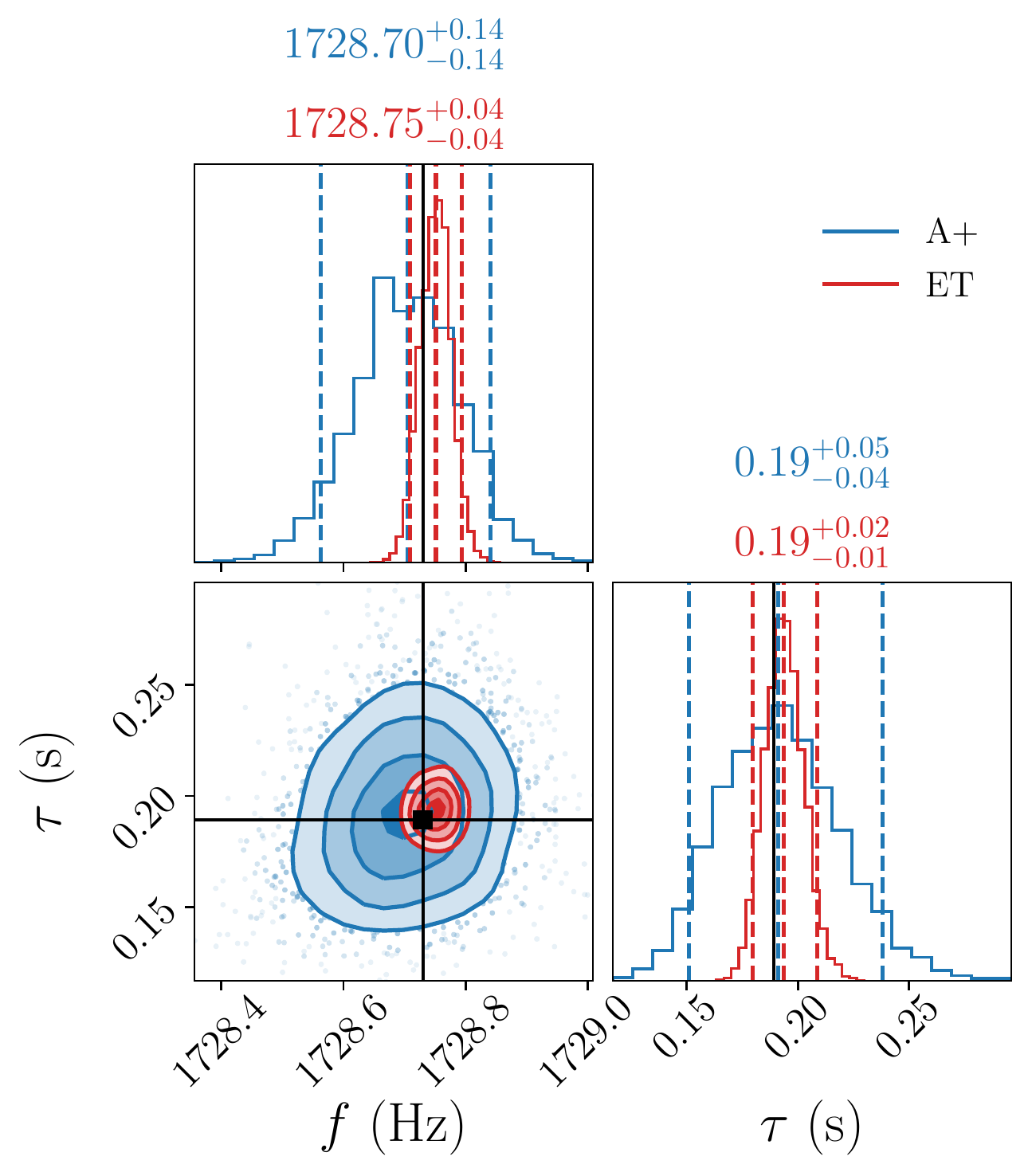}
    \caption{The joint corner and marginalized posterior distribution of the recovered ($f,\tau$) for an injected event corresponding to  one of the Vela glitches with $\frac{\Delta \nu}{\nu}=3.1\times 10^{-6}$ in $\rm A+$ (blue in color) and ET (red) GW detector configurations. The black lines display the injected values. The 90\% SCIs are also shown with dashed lines. The errors shown on top correspond to the 90\% SCI.}
    \label{fig:f_tau_oneevent}
\end{figure}

We consider a GW event corresponding to the glitch of Vela pulsar with $\frac{\Delta \nu}{\nu}=3.1\times 10^{-6}$ in $\rm A+$ and ET configuration. The randomly assigned mass to the Vela pulsar in the process described in Section ~\ref{sec:methodology} is $\sim1.84 M_{\odot}$. Further, we assign the other required parameters corresponding to 1.84$M_{\odot}$ NS from assumed PCSB0 EOS. We display the recovered joint posterior of f-mode frequency ($f$) and damping time ($\tau$) in ~\cref{fig:f_tau_oneevent}. From ~\cref{fig:f_tau_oneevent}, one can conclude that the mode frequency can be recovered within $\leq 1\%$ (within 90\% SCI) in both $\rm A+$ and ET configuration. However, in $\rm A+$ detector configuration, the recovered damping time $\tau$ can have $\sim 27\%$ uncertainty  (within 90\% SCI). In contrast, the error in the measurement on damping time reduced to $~10\%$ in the ET configuration\footnote{For a considered glitch event, the uncertainties on $f$ and $\tau$ might depend upon the assigned injection frequency, damping time, and the sensitivity of the GW detector~\citep{Kokkotas2001}. We have checked this by varying the assigned mass to the pulsar glitches and found that in every case, $f$ is well recovered within $\sim 1\%$ (within 90\% SCI). However, the damping time can have uncertainty (within 90\% SCI) $\sim 20\% - 50\%$ and $\sim 8\% - 20\%$ in $\rm A+$  and ET, respectively.}.  The resulting uncertainties on $f$ and $\tau$ agree with the errors predicted by ~\citet{Kokkotas2001} resulting from the Fisher matrix analysis. Hereafter, the uncertainties on all the physical parameters, including nuclear parameters, EOS properties, and  NS properties, are quoted within 90\% SCIs (unless mentioned otherwise).

\begin{figure*}
    \centering
    \includegraphics[width=\linewidth]{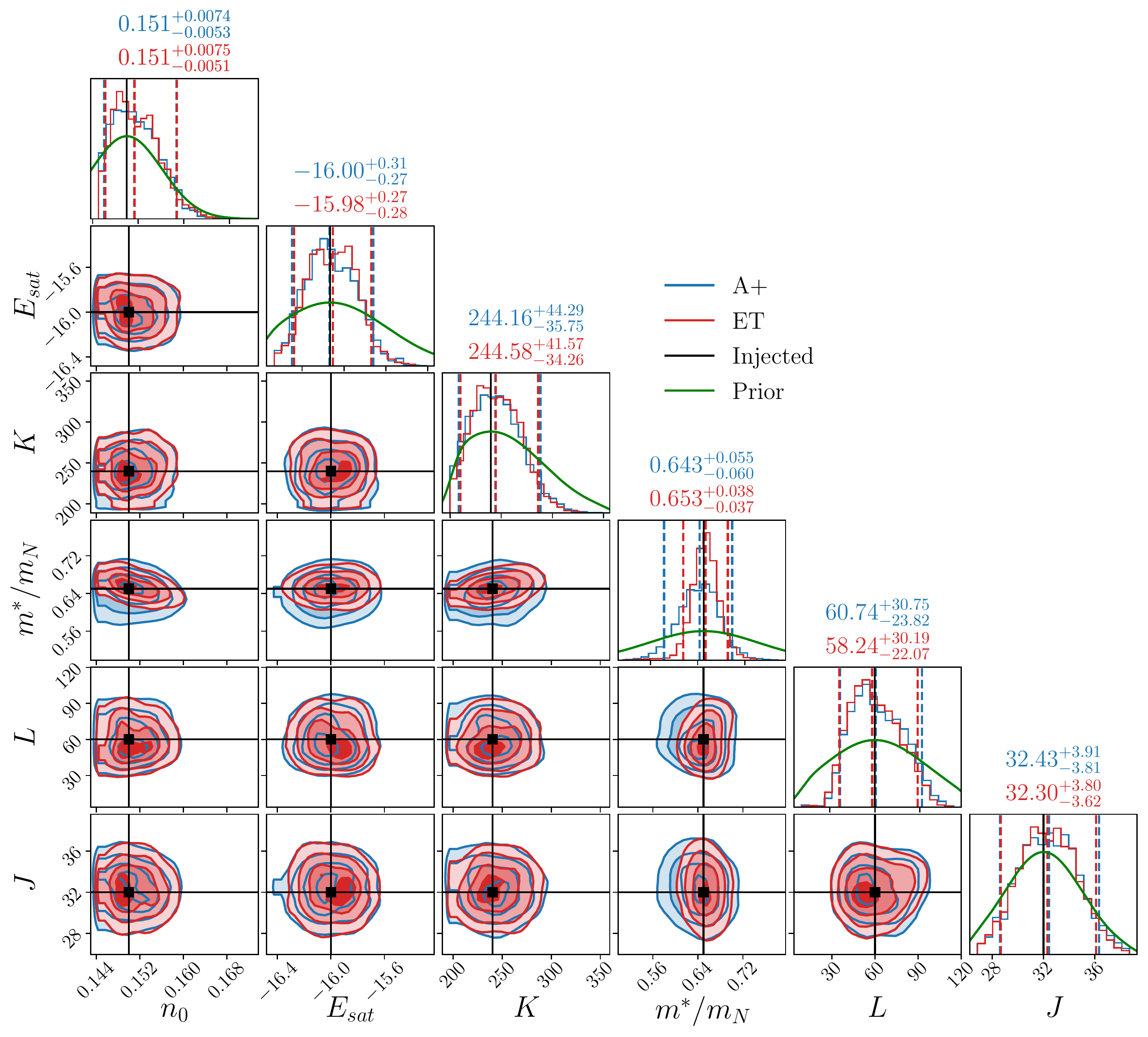}
    \caption{Joint distribution of the posteriors of the nuclear parameters recovered  from the   posterior of ($f,\tau$) (see \cref{fig:f_tau_oneevent}) for the considered event in A+ (blue line) and ET (red lines). The black lines show the injected values. The median and 90\% SCIs are also shown in the respective marginalized one-dimensional posterior histograms. The considered priors are  also shown in green color.}
    \label{fig:nuclear_Velaglitch}
\end{figure*}

For the considered GW event in A+ and ET, the resulting posteriors of the nuclear parameters and prior distributions are displayed in ~\cref{fig:nuclear_Velaglitch} and tabulated in ~\cref{tab:prior_posterior}. From, ~\cref{fig:nuclear_Velaglitch}, one can conclude that from detecting a single event, the nucleon effective mass ($m^*/m_N$)  can be constrained up to ~10\% and 5\% in A+ and ET, respectively. Among the other nuclear parameters, the estimation of $K$ and $L$ slightly improved compared to the prior used. It has been shown that for the considered RMF model, $m^*$ primarily controls the softness of the EOS and hence shows strong correlations with the mode frequency and damping time ~\citep{Jaiswal,Pradhan2022,Pradhan2023DAE}, which explains that the detection of f-mode helps us constrain $m^*$ more precisely compared to the other nuclear parameters for the considered RMF model. 

\begin{figure*}
\centering 

\subfigure[]{%
  \includegraphics[width=0.35\textwidth]{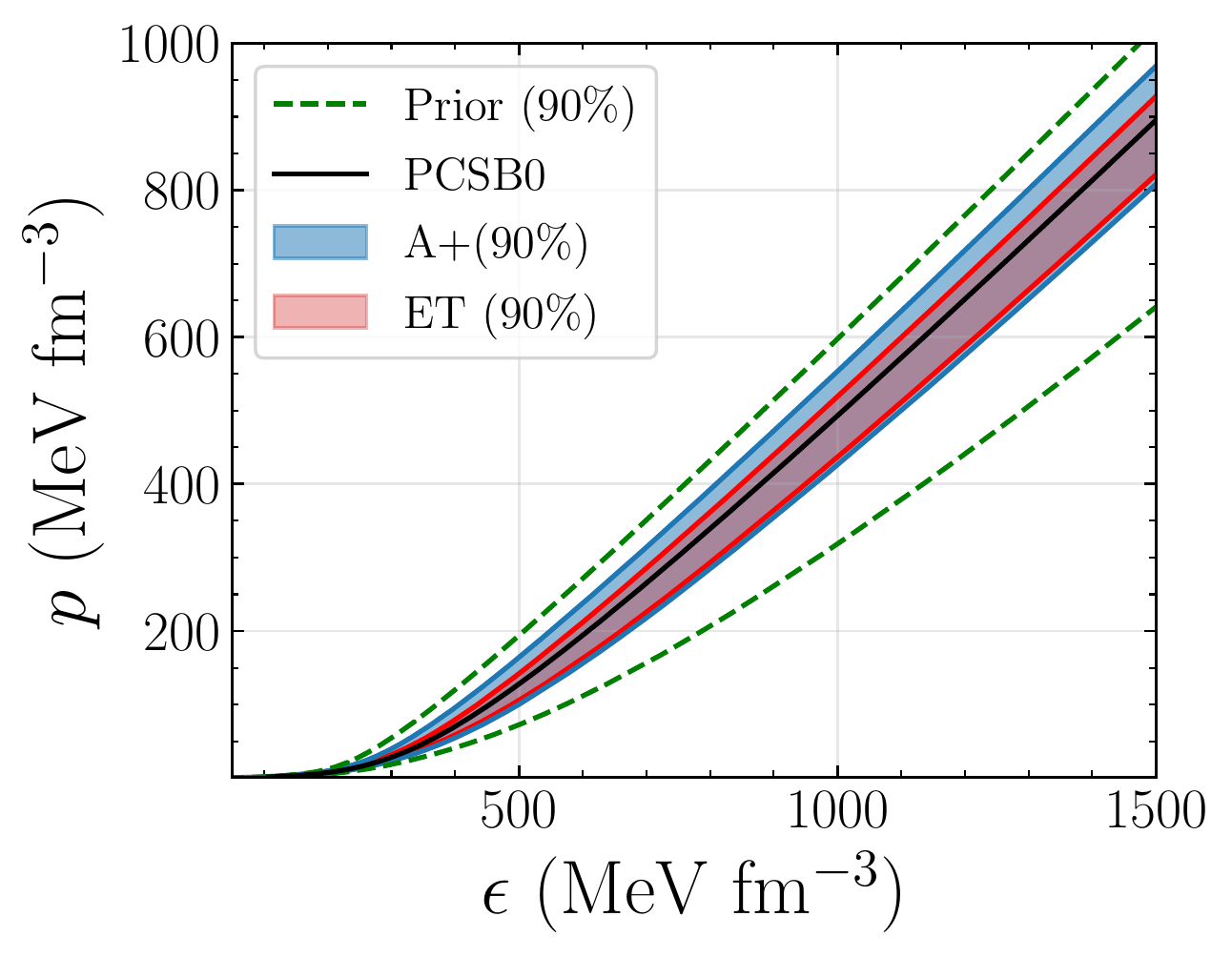}%
  \label{subfig:1event_EOS}%
}
\subfigure[]{%
  \includegraphics[width=0.33\textwidth]{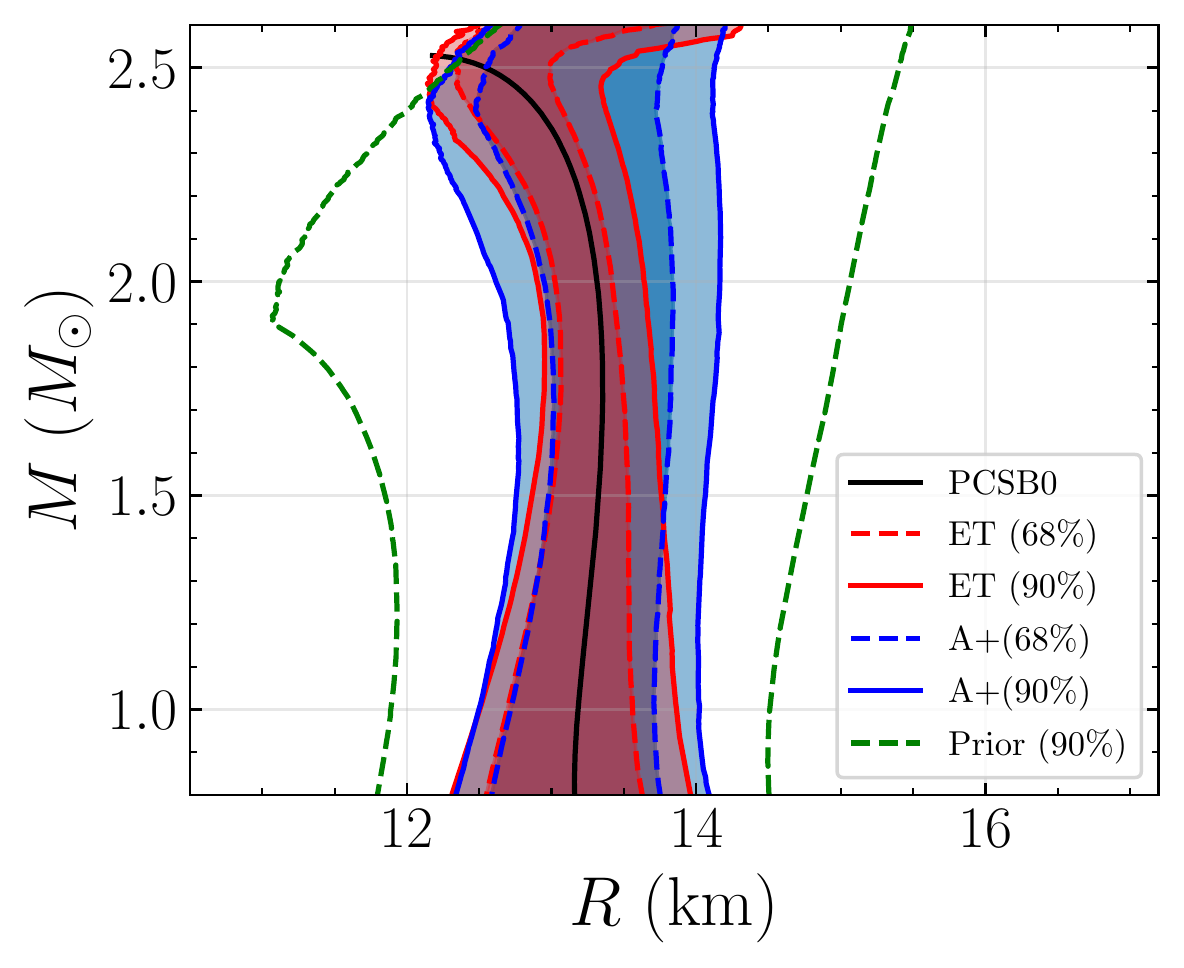}%
  \label{subfig:1event_mr}%
}\\
\subfigure[]{%
  \includegraphics[width=0.35\textwidth]{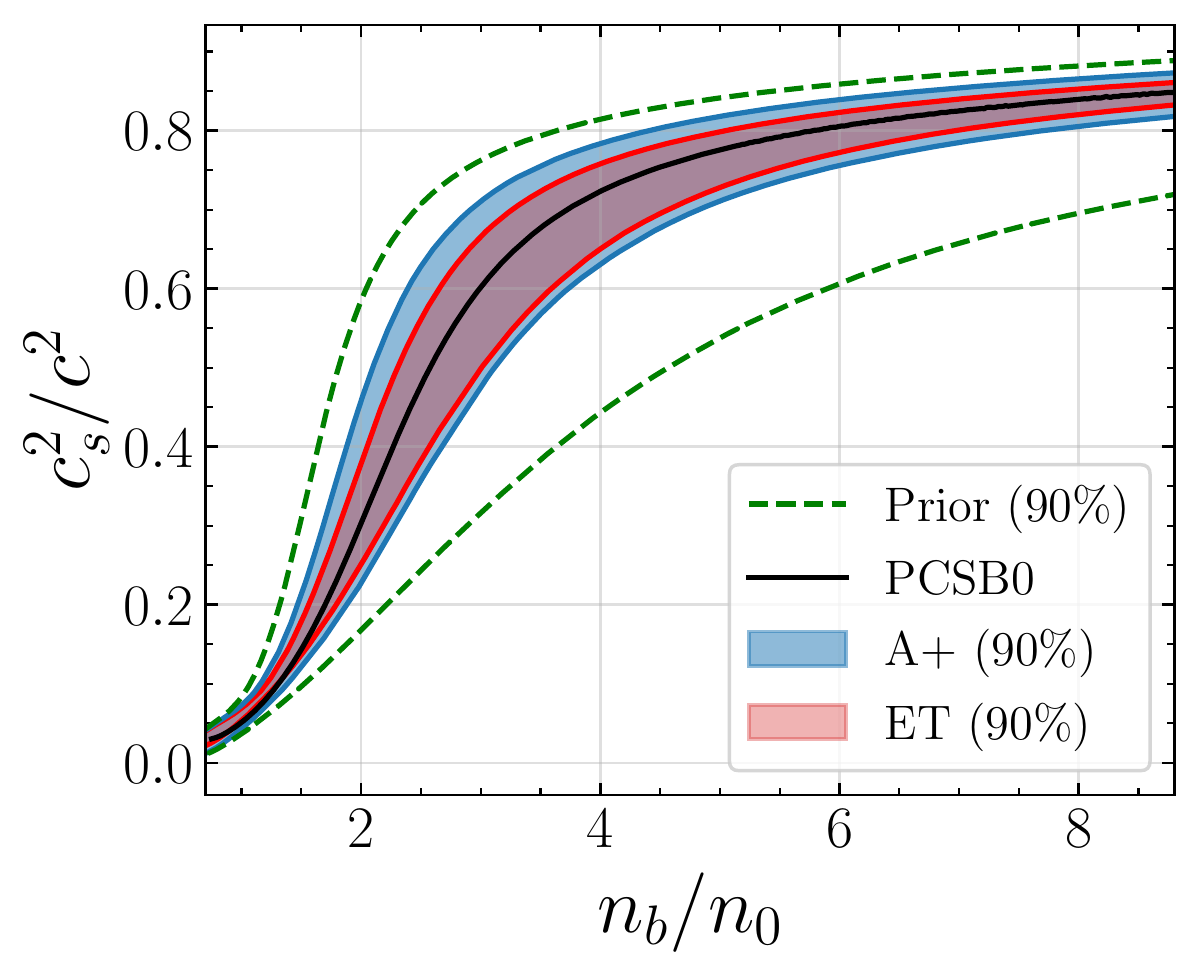}%
  \label{subfig:1event_cs}%
}
\subfigure[]{%
  \includegraphics[width=0.35\textwidth]{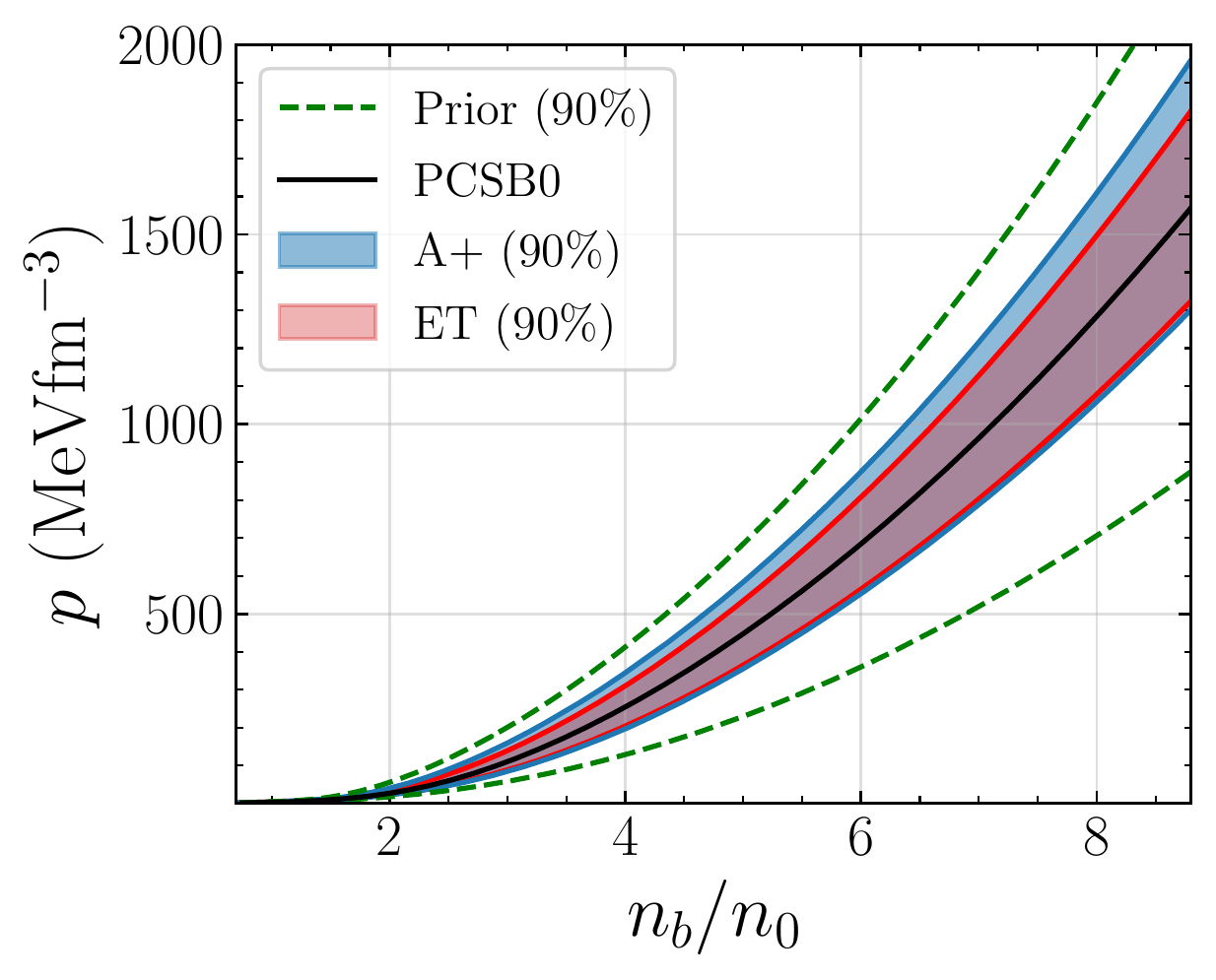}%
  \label{subfig:1event_nbp}%
}
\caption{We plot 90\% symmetric SCI of    (a) the pressure as a function of energy density, (b) the radius as a function of mass,  (c) the squared speed of sound as a function of density and (d) pressure as a function of density,  reconstructed using the posteriors of nuclear parameters corresponding to the single f-mode event displayed in~\cref{fig:nuclear_Velaglitch}.  To avoid clutter, we show both the 68\% (bounded by dashed lines)  and 90\% (bounded by solid lines) SCIs only for $M-R$ relations.}
\label{fig:EOS_properties_1event}
\end{figure*}

Although the uncertainty in the other nuclear parameters improved marginally, the tighter constraint on $m^*$ will help us improve the tension in the EOS measurements and NS properties (as for the RMF model considered, $m^*$ mainly controls the stiffness of the EOS). For the obtained posteriors, we reconstruct the EOSs and other NS observable and tabulate some of the properties of the EOS and NS in ~\cref{tab:EOS_NS_properties}. We display the uncertainty in the measurement of EOS, mass-radius, squared speed of sound ($c_s^2$) as a function of number density, and pressure as a function of number density in ~\cref{subfig:1event_EOS,subfig:1event_mr,subfig:1event_cs,subfig:1event_nbp}, respectively.

\begin{figure*}
\centering 

\subfigure[]{%
  \includegraphics[width=0.42\textwidth]{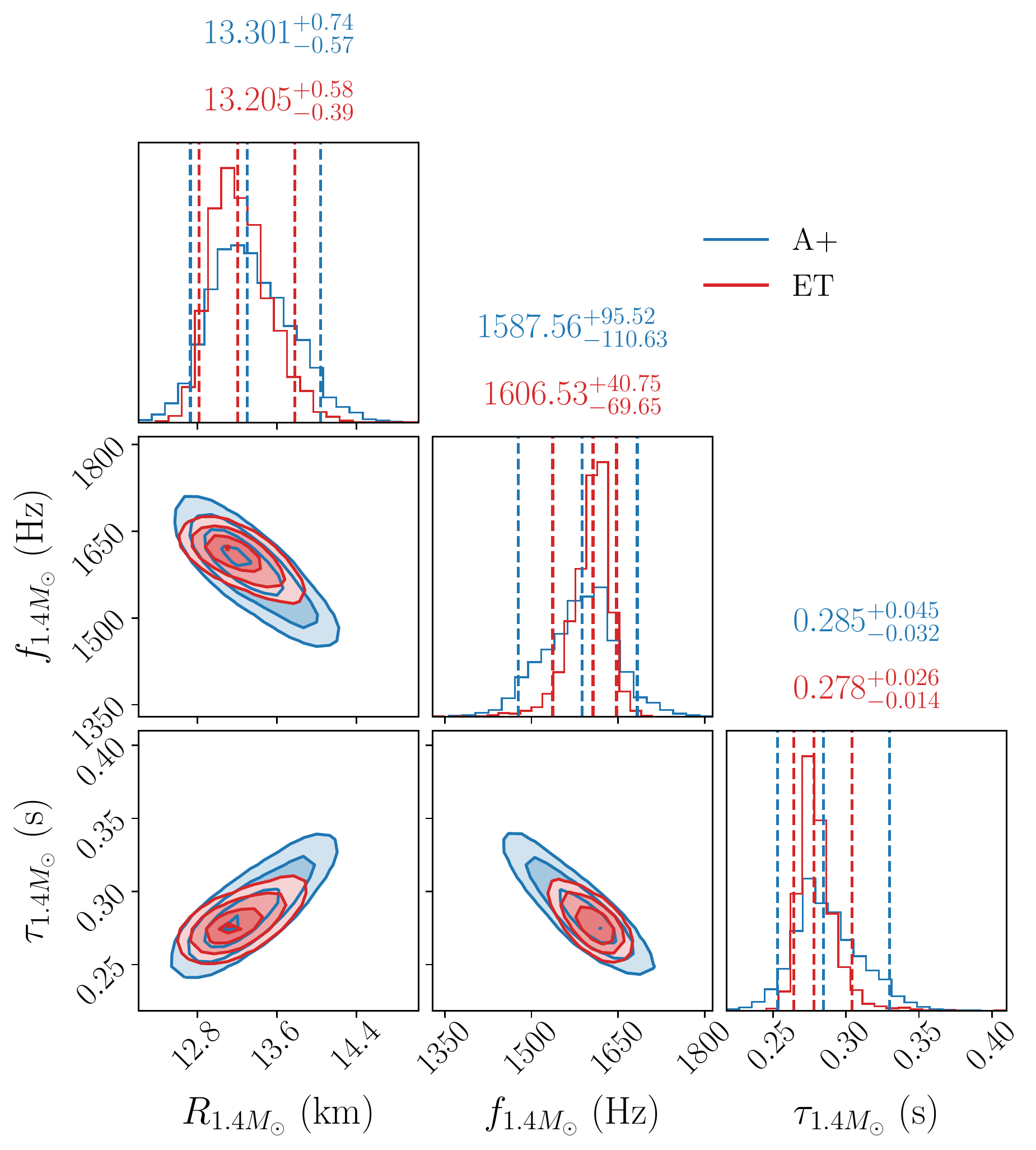}%
  \label{fig:NS_properties_1event}%
}
\subfigure[]{%
  \includegraphics[width=0.42\textwidth]{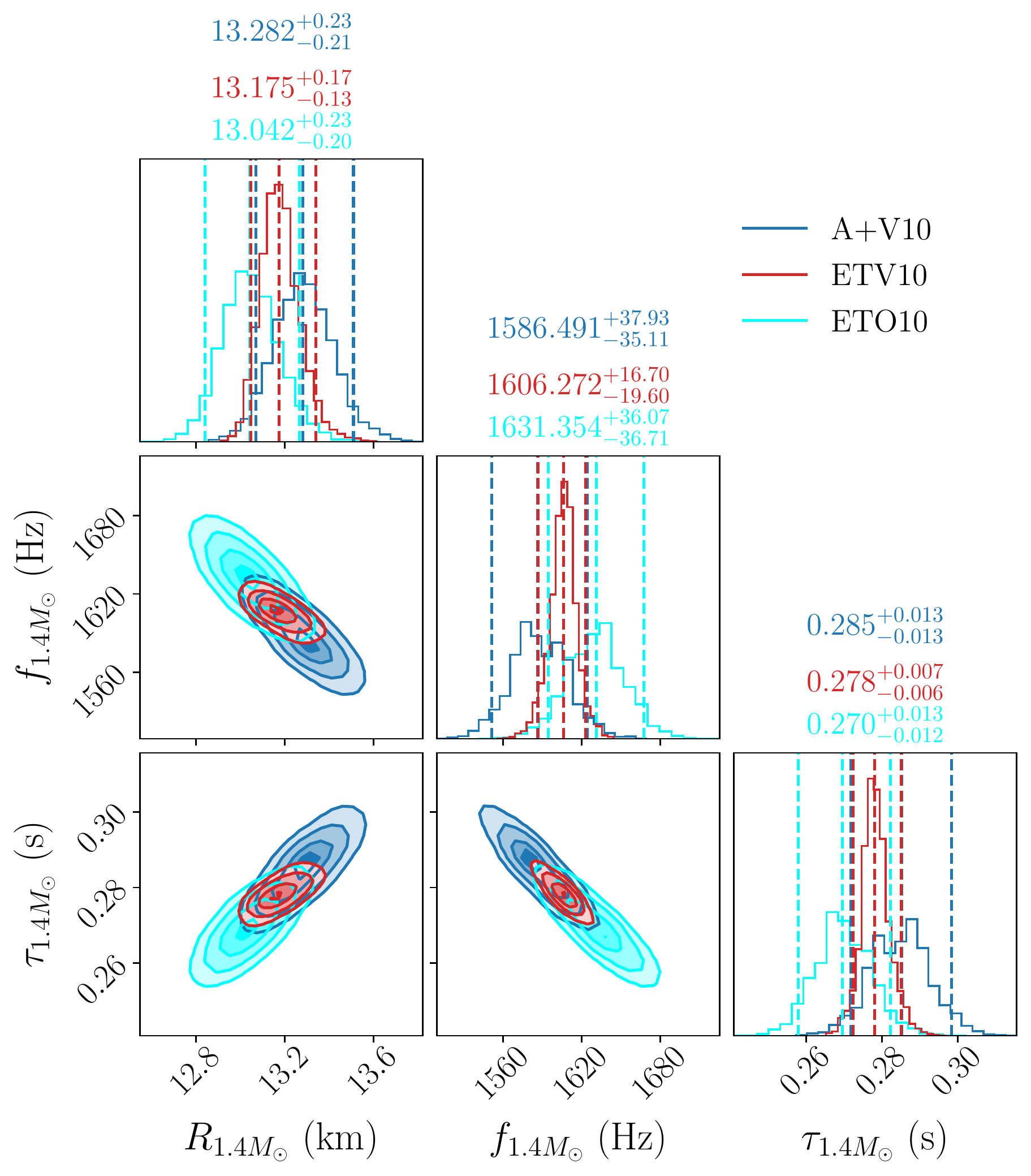}
    \label{fig:NS_properties_10event}
}
\caption{(a) The posterior distribution of $R_{1.4M_{\odot}}$, $f_{1.4M_{\odot}}$ and $\tau_{1.4M_{\odot}}$ resulting from the detection of a single event in A+ (blue) and ET (red). (b) Similar to ~\cref{fig:NS_properties_1event} but with consideration of  10 random events  resulting from  Vela with SNR $\geq 8$  in A+ (blue) and ET (red) and the posterior resulting form consideration of 10 random events in ET with pulsars other than Vela.   }

\end{figure*}

We display the distribution of  radius ($R_{1.4M_{\odot}}$),  frequency ($f_{1.4M_{\odot}}$), and damping time ($\tau_{1.4M_{\odot}}$) of a canonical ${1.4M_{\odot}}$ NS (reconstructed from the posterior displayed in ~\cref{fig:nuclear_Velaglitch} ) in  ~\cref{fig:NS_properties_1event} and tabulated in ~\cref{tab:EOS_NS_properties}.  In A+, the $R_{1.4M_{\odot}}$, $f_{1.4M_{\odot}}$, and $\tau_{1.4M_{\odot}}$ can be measured with in  ~6\%, 7\%, and  18\%, respectively. With ET,  of $R_{1.4M_{\odot}}$, $f_{1.4M_{\odot}}$, and $\tau_{1.4M_{\odot}}$ can be measured within  ~4\%, 4\%, and  9\%, respectively. 

 From ~\cref{fig:EOS_properties_1event}, one can conclude that the injected EOS and its properties (i.e., the corresponding $M-R$ relation, $c_s^2$ and $p$) are well recovered in both A+ and ET. However, the uncertainty in measurements improved significantly with the ET configuration compared to A+, which is expected with the increasing sensitivity of ET. We discuss the impact of consideration of other existing astrophysical observations along with the future f-mode detection on the estimation of nuclear parameters and EOS uncertainties in Appendix ~\ref{app:Astro_and_fmode}. However, for completeness, we have tabulated the posteriors of the nuclear parameters in ~\cref{tab:prior_posterior} and also tabulated the NS and EOS properties in ~\cref{tab:EOS_NS_properties}.

\begin{table*}
    \centering\small\setlength\tabcolsep{.22em}
    \begin{tabular}{|c|c|c|c|c|c|c|}
    \hline
    \hline
        Parameter& Injection & Prior & \multicolumn{4}{c|}{Posterior}   \\
         \cline{4-7}
         & & &  1 Event in A+  & Astro + 1 Event in A+  &  A+V10 & ETO10 \\
          & (PCSB0)  & &  (1 Event in  ET ) & (Astro + 1 Event in  ET )& (ETV10) &\\
          \hline
          \hline
         $n_0 \rm{[fm^{-3}]}$&0.15  &$\mathcal{N} (0.15,0.006) \ T[0.14,0.18] $ & $0.151^{+0.0074}_{-0.0053}$ & $0.151^{+0.0035}_{-0.0035}$ &$0.150^{+0.0025}_{-0.0019}$ &$0.150^{+0.0031}_{-0.0023}$\\
         & & & $(0.151^{+0.0075}_{-0.0051})$ & $(0.150^{+0.0038}_{-0.0034})$ &$(0.150^{+0.0020}_{-0.0017})$ &\\
         \hline
         $E_{sat}\rm{[MeV]}$&-16.0   &$\mathcal{N} (-16,0.40)\  T[-16.5,-15] $ &$-16.00^{+0.31}_{-0.27}$ & $-16.01^{+0.22}_{-0.19}$  & $-15.99^{+0.10}_{-0.10}$ & $-15.99^{+0.11}_{-0.12}$ \\
         & & & $(-15.98^{+0.27}_{-0.28})$ & $(-15.99^{+0.21}_{-0.20})$ &$(-16^{+0.10}_{-0.10})$  & \\
         \hline 
         $K\rm{[MeV]}$&240 &$\mathcal{N} (240,50)\  T[200,355] $ & $244^{+44}_{-36}$ & $243^{+27}_{-23}$ & $236^{+21}_{-13}$&$237^{+20}_{-16}$\\
         & & &$(245^{+42}_{-34})$ & $(241^{+28}_{-23})$ & ($236^{+20}_{-13}$) & \\
         \hline
         $m^*/m_N$ &0.65&$\mathcal{N} (0.65,0.09)\  T[0.4,0.9] $ & $0.643^{+0.055}_{-0.060}$ & $0.673^{+0.035}_{-0.036}$ &$0.640^{+0.021}_{-0.020}$ &$0.662^{+0.021}_{-0.022}$\\
         & & & $(0.653^{+0.038}_{-0.037})$ & $(0.655^{+0.027}_{-0.028})$ & $(0.648^{+0.011}_{-0.010})$ & \\
         \hline 
         $L\rm{[MeV]}$&60 &$\mathcal{N} (60,35)\  T[1,140] $ & $61^{+31}_{-24}$ & $55^{+20}_{-16}$ &$56^{+11}_{-8}$ &$52^{+11}_{-9}$ \\
         & & & $(58^{+30}_{-22})$ & $(56^{+20}_{-16})$ &($54^{+11}_{-7}$)  & \\
         \hline
         $J\rm{[MeV]}$&32 &$\mathcal{N} (32,3)\  T[26,39] $ & $32.43^{+3.91}_{-3.81}$ & $32.22^{+1.98}_{-2.04}$ &$32.22^{+1.20}_{-1.17}$ &$31.71^{+1.63}_{-1.43}$\\
          & & & $(32.30^{+3.8}_{-3.62})$ & $(32.23^{+1.97}_{-2.06})$ & $(32.49^{+1.22}_{-1.36}$) & \\
         \hline
         
        \hline \hline
    \end{tabular}
    \caption{The median and 90\% symmetric credible interval of the recovered  posterior of  nuclear parameters resulting from different scenarios involving  GW observation from NS f-mode. The true parameters of PCSB0 EOS are also tabulated in injection column. The priors considered are  truncated Gaussians with mean at the parameters of PCSB0 EOS and deviations are inspired from ~\citet{Margueron2018}. We also truncated the distribution in a range such that it includes  the minimum and maximum ranges resulting from the phenomenological models as given in ~\citet{Margueron2018}.  }
    \label{tab:prior_posterior}
\end{table*}

\begin{table*}
    \centering\small\setlength\tabcolsep{.28em}
    \begin{tabular}{|c| c| c|c |c|}
    \hline
    \hline
        Parameters &  \multicolumn{4}{c|}{Posterior}  \\
         \hline
          &  1 Event in A+  & Astro + 1 Event in A+  & A+V10  & ETO10 \\
           &  (1 Event in  ET ) & (Astro + 1 Event in  ET )& (ETV10) &\\
          \hline \hline
           \textbf{EOS properties} & & & & \\
          \hline \hline
         $p(2n_0)\ \rm [MeV\  fm^{-3}]$ &\onePAplusatftwon &\onePAplusastroattwon &\tenPAplusatftwon & \allPETattwon\\
          & (\onePETattwon) & (\onePETastroattwon) &(\tenPETattwon)& \\
         \hline
         $p(5n_0)\ \rm [MeV\  fm^{-3}]$ &\onePAplusatfiven & \onePAplusastroatfiven&\tenPAplusatfiven & \allPETatfiven\\
          & (\onePETatfiven) &(\onePETastroatfiven) &(\tenPETatfiven)& \\
         \hline
         
         $c_s^2(2n_0)\  [c^2]$ &\onecsAplusatftwon & \onecsAplusastroattwon&\tencsAplusatftwon & \allcsETattwon\\
          & (\onecsETattwon) &(\onecsETastroatfiven) &(\tencsETattwon)& \\
         \hline
         $c_s^2(5n_0)\  [c^2]$ &\onecsAplusatfiven & \onecsAplusastroatfiven&\tencsAplusatfiven & \allcsETatfiven\\
          & (\onecsETatfiven) & (\onecsETastroatfiven)&(\tencsETatfiven)& \\
         \hline \hline
         \textbf{NS Properties} & & & &\\
         \hline \hline
         $R_{1.4M_{\odot}}\ \rm [km]$ & $13.30^{+0.74}_{-0.57}$ & $13.03^{+0.44}_{-0.39}$& $13.28^{+0.23}_{-0.21}$&$13.04^{+0.23}_{-0.20}$ \\
         &($13.20^{+0.58}_{-0.39}$)& ($13.10^{+0.34}_{-0.32}$)&($13.18^{+0.17}_{-0.13}$) &\\
         \hline
         $f_{1.4M_{\odot}}\ \rm [Hz]$ &$1588^{+96}_{-111}$  &$1641^{+67}_{-68}$ &$1587^{+38}_{-35}$ & $1631^{+36}_{-37}$\\
          &($1607^{+41}_{-70}$)  &($1626^{+50}_{-47}$) & ($1606^{+17}_{-20}$) & \\
         \hline
         $\tau_{1.4M_{\odot}}\ \rm [s]$  &$0.285^{+0.045}_{-0.032}$  & $0.266^{+0.024}_{-0.021}$ &$0.285^{+0.013}_{-0.013}$ &$0.27^{+0.013}_{-0.012}$ \\
          &($0.278^{+0.026}_{-0.014}$)  & ($0.271^{+0.017}_{-0.016}$)&($0.278^{+0.007}_{-0.006}$)& \\
        \hline \hline
    \end{tabular}
    \caption{The median and 90\% symmetric credible interval of some of the EOS and NS properties resulting from the recovered posteriors of nuclear parameters in different scenarios. }
    \label{tab:EOS_NS_properties}
\end{table*}

\subsection{Multiple events}

It has been discussed in ~\citet{Ho2020} (also shown here in ~\cref{fig:snr}), there could be many events originating from different pulsars that could be detected in the future. In A+ configuration, f-mode GWs can be detected from  the Vela pulsar under the assumed scenario. However, in ET, sources from different pulsars  could have SNR $\geq 10$, indicating many potential sources. We consider two different scenarios to see the impact of multiple events on the posterior of nuclear parameters or the EOS measurement:
\begin{itemize}
\item First, we consider ten events from Vela with various glitch energy  having SNR $\geq 8$ in A+  configuration and analyze the effect: we label this as A+V10. We consider the same ten events in ET (for the considered sources, SNR can be $\geq 20$), which helps us compare the posteriors in two different GW detector configurations: we label this as ETV10. 
\item In the second scenario, with ET, we consider 10 random f-mode GW events of different pulsars other than Vela with SNR $\geq 8$ and investigate the impact. This scenario is labeled as ETO10.
\end{itemize}

\begin{figure*}
    \centering
    \includegraphics[width=\linewidth]{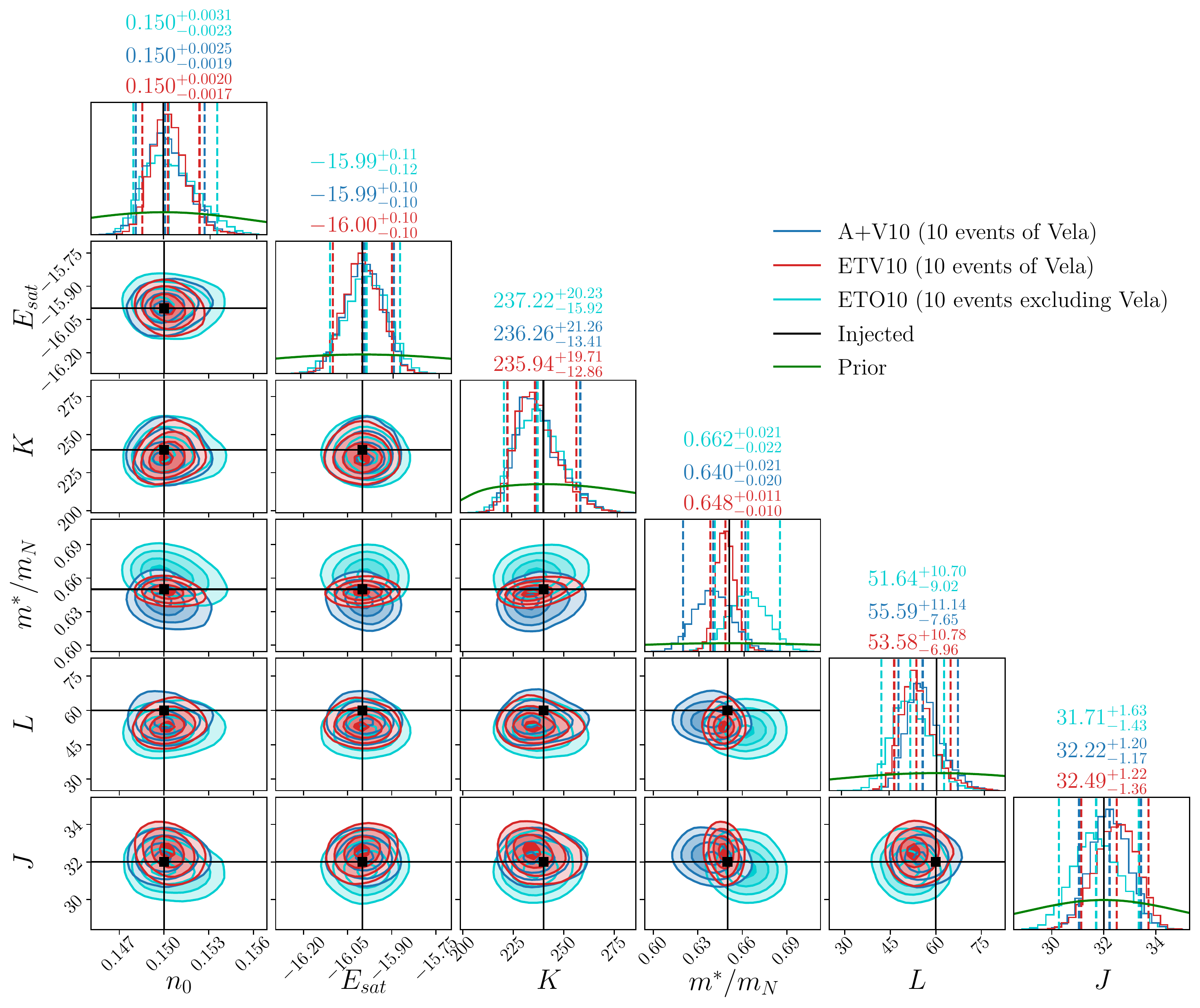}
    \caption{Similar to ~\cref{fig:nuclear_Velaglitch} but with consideration of  10 random events with SNR $\geq 8$  in A+ (blue line) and ET (red lines) corresponding to f-mode events from Vela pulsar.  We have also shown the distributions of nuclear parameters resulting from considering  10 random events in ET, excluding the Vela (cyan color).}
    \label{fig:nuclear_10_events}
\end{figure*}

For ten events in A+ and ET, the posterior distributions of recovered nuclear parameters are displayed in ~\cref{fig:nuclear_10_events}. In both A+ and ET, one can find from ~\cref{fig:nuclear_10_events} that the distribution of Gaussian priors appear to be flat in the range of the posteriors of the parameters indicating tighter constraints on the posteriors of the nuclear parameters. All nuclear parameters are well constrained in both ET and A+ with ten events. The posteriors of the nuclear parameters are tabulated in ~\cref{tab:prior_posterior}. In A+, within 90\% SCI, $m^*$ can be constrained within 3\%. The uncertainty in $m^*$ improves in ET and  can be constrained within 2\%.
{It is interesting to note that by combining multiple events in A+ ( in ET), within 90\% SCI the incompressibility ($K$) and the slope of symmetry energy ($L$) can be constrained  to $\sim$10\% and $\sim$ 20\%, respectively.}

\begin{figure*}
\centering 

\subfigure[]{%
  \includegraphics[width=0.35\textwidth]{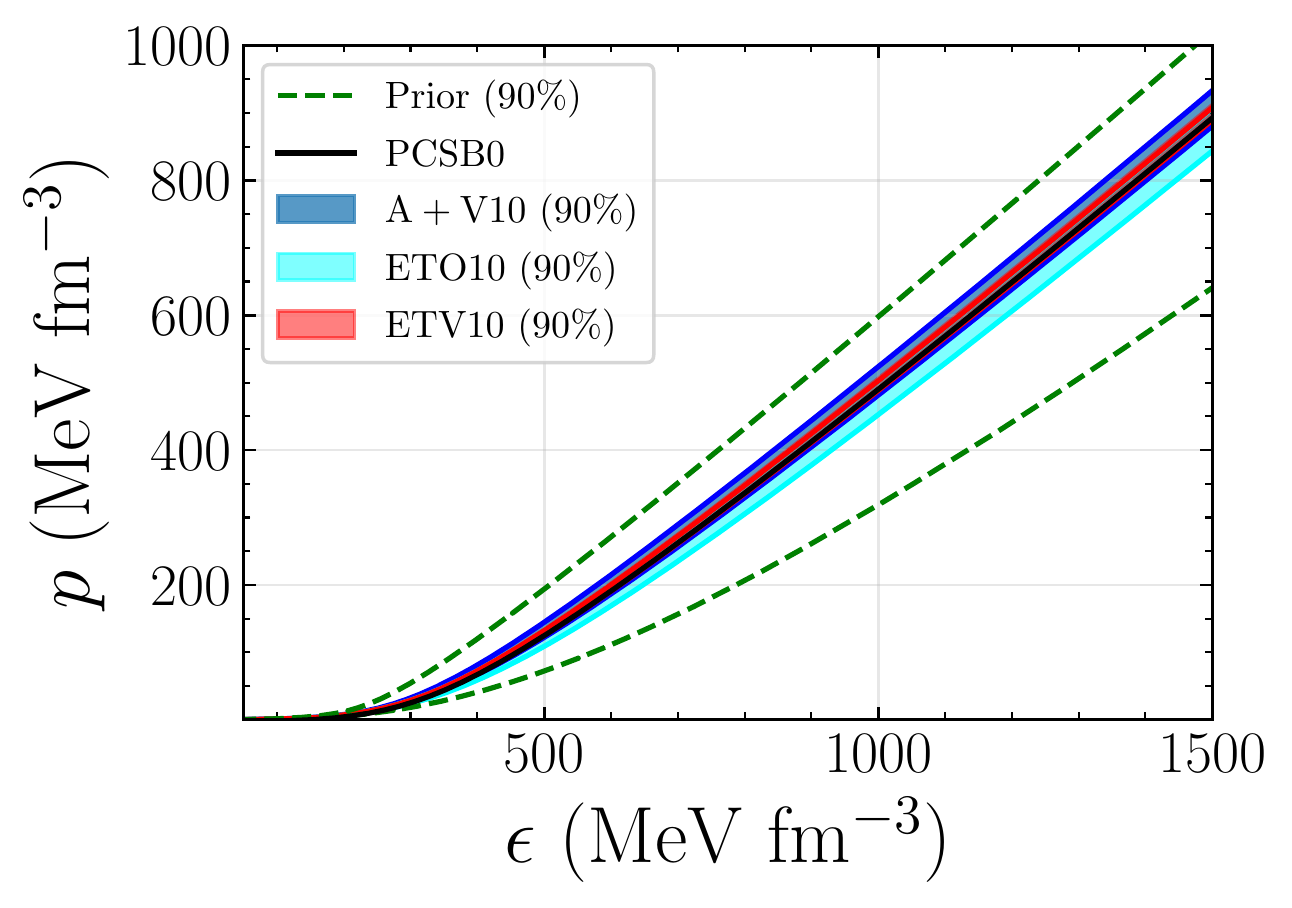}%
  \label{subfig:10event_EOS}%
}
\subfigure[]{%
  \includegraphics[width=0.33\textwidth]{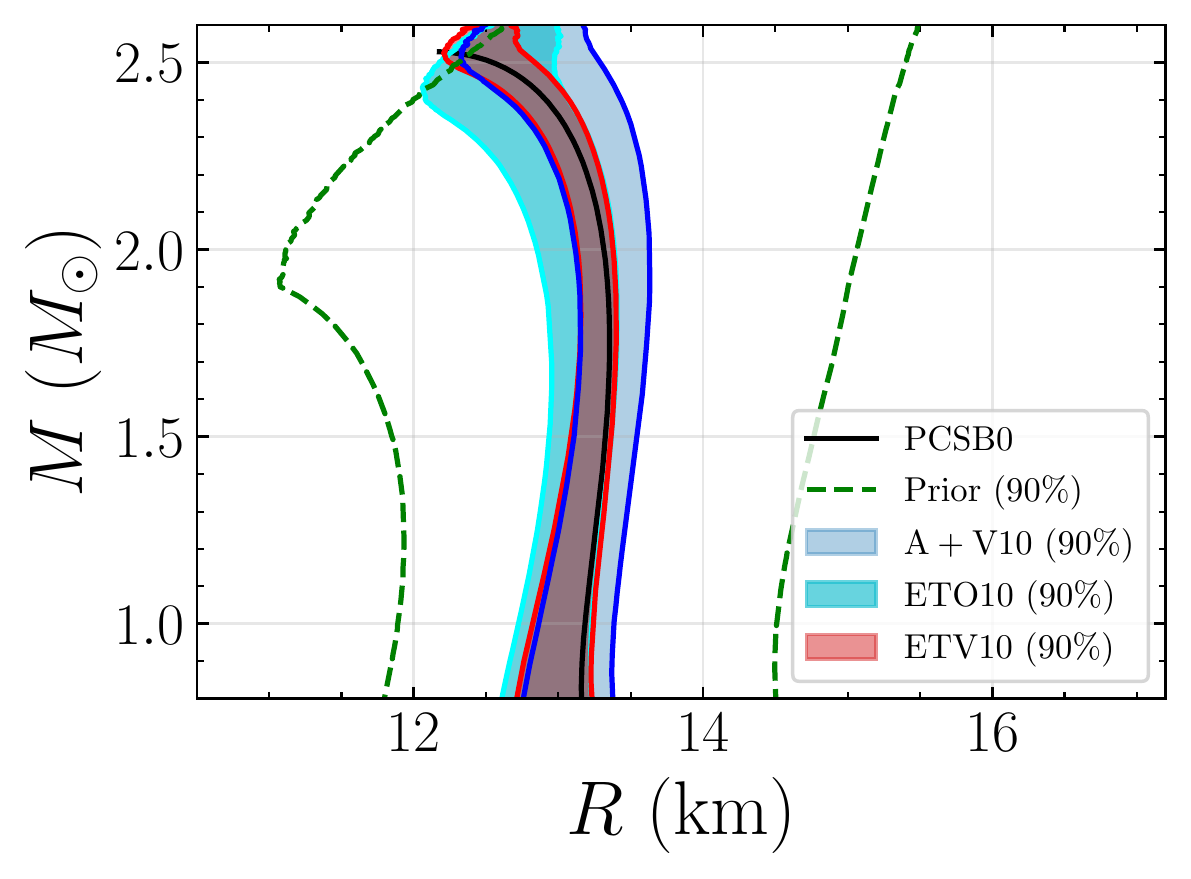}%
  \label{subfig:10event_mr}%
}\\
\subfigure[]{%
  \includegraphics[width=0.33\textwidth]{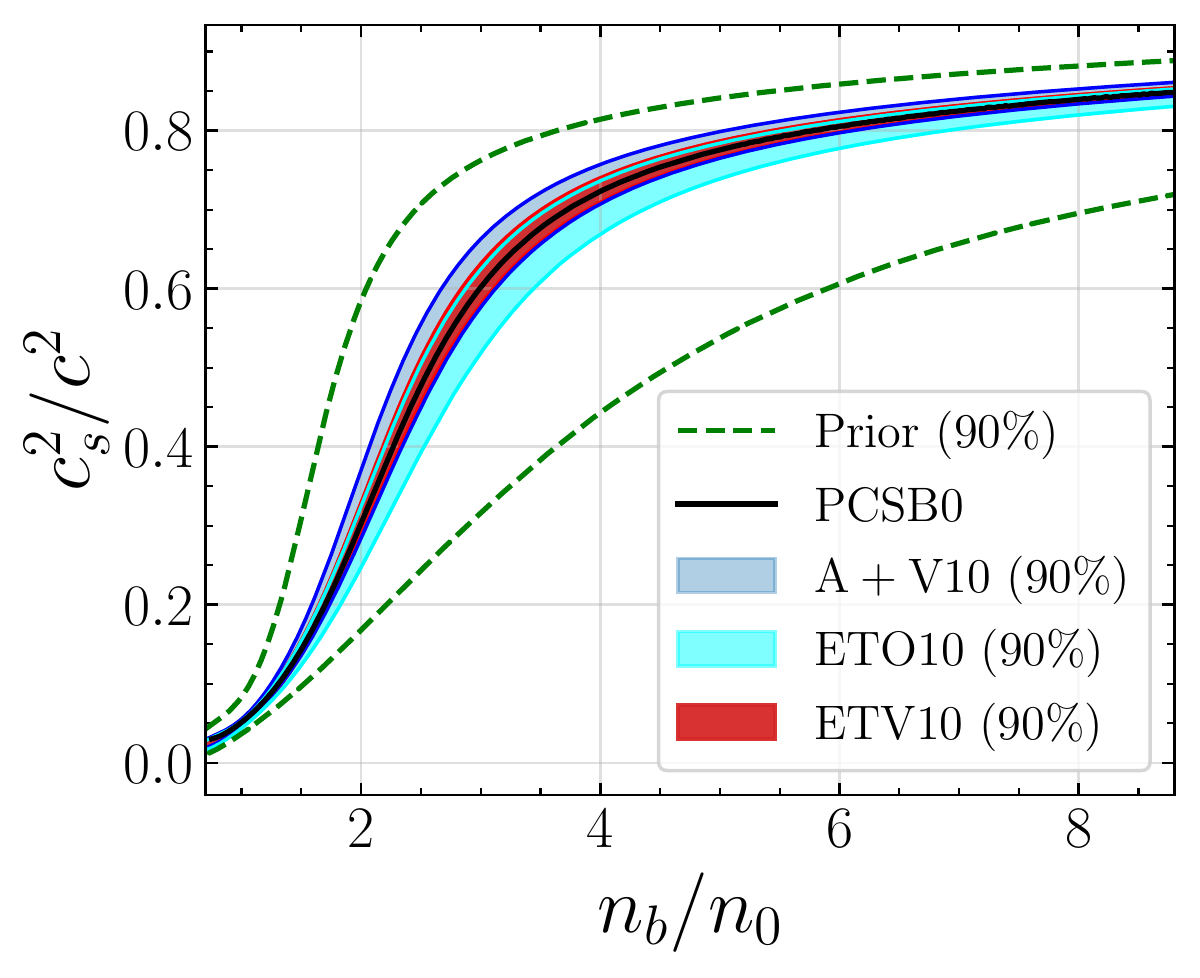}%
  \label{subfig:10event_cs}%
}
\subfigure[]{%
  \includegraphics[width=0.35\textwidth]{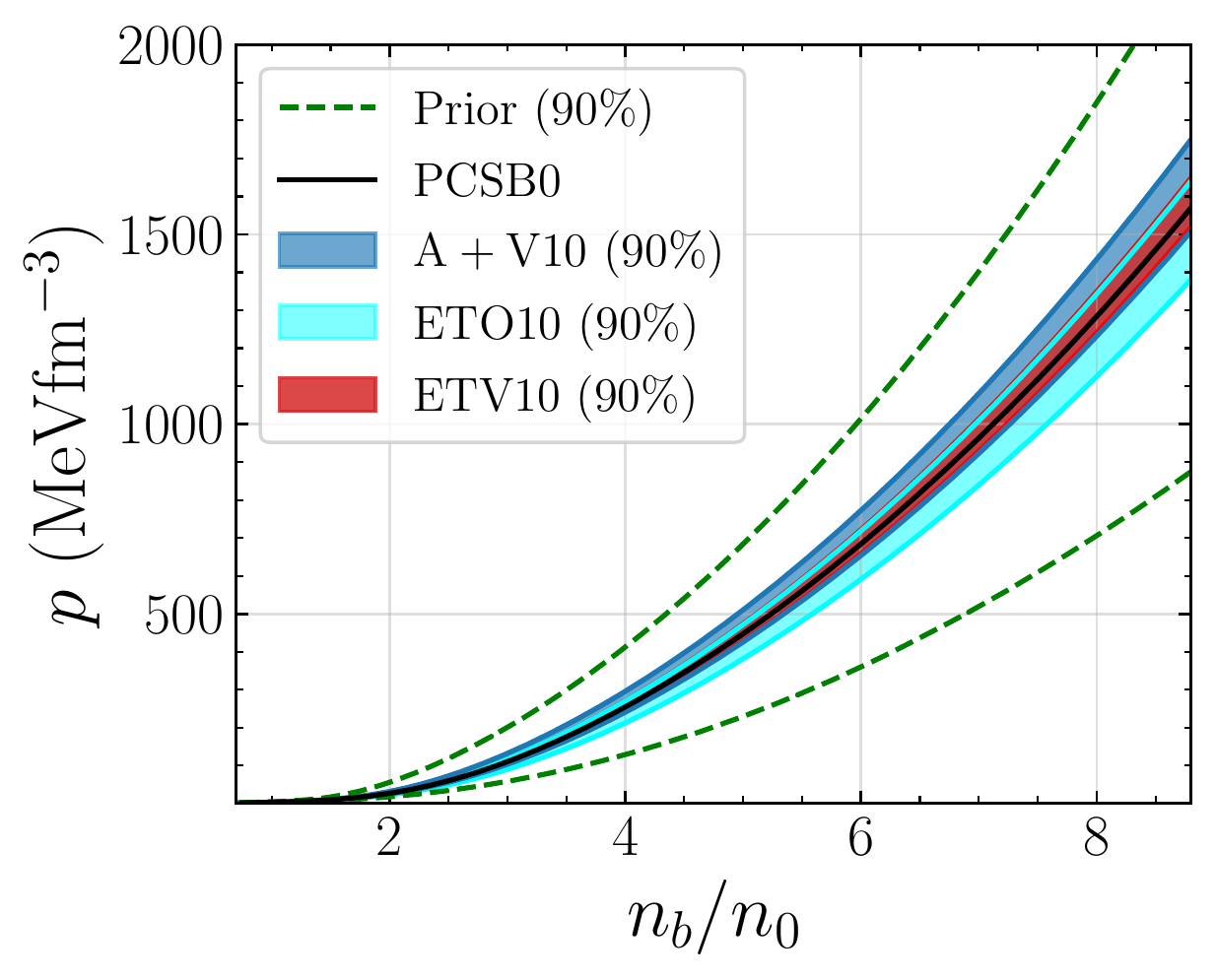}%
  \label{subfig:10event_nbp}%
}
\caption{Similar to ~\cref{fig:EOS_properties_1event} but with consideration of  10 random events with SNR $\geq 8$  in A+ (blue band) and ET (red band) corresponding to glitches of Vela pulsar. We have also displayed the uncertainties resulting from 10 random events in  ET, excluding the glitches from Vela pulsar (cyan band)}
\label{fig:EOS_properties_10event}
\end{figure*}

The uncertainties in the measurements of EOS, $M-R$, $c_s^2$ and $p$, reconstructed from the recovered posteriors of nuclear parameters with ten events, are displayed in  ~\cref{subfig:10event_EOS,subfig:10event_mr,subfig:10event_cs,subfig:10event_nbp} respectively. We tabulate the  EOS and NS properties recovered with ten events in ~\cref{tab:EOS_NS_properties}. Although all the nuclear parameters except $m^*$ are constrained within similar accuracy in both A+ and ET, the tighter posterior of $m^*$ in ET compared to A+ improves the uncertainty of the measurements of the EOS and NS properties (see, ~\cref{fig:EOS_properties_10event} or ~\cref{tab:EOS_NS_properties}). In A+ (ET),  $p$ and  $c_s^2$ at $5n_0$ can be estimated within $\sim 10\%$ (5\%) and $\sim 2\%$ (1\%). We display the distribution of the posteriors for  recovered $R_{1.4M_{\odot}}$, $f_{1.4M_{\odot}}$, and $\tau_{1.4M_{\odot}}$ in ~\cref{fig:NS_properties_10event} and tabulate the median and 90\% SCIs in ~\cref{tab:EOS_NS_properties}. $f_{1.4M_{\odot}}$  can be constrained within 40 Hz and 20Hz with A+ and ET, respectively (see \cref{fig:NS_properties_10event}). The uncertainty in the measurement of $\tau_{1.4M_{\odot}}$ improved significantly compared to a single event. In A+ (ET),  $\tau_{1.4M_{\odot}}$ can be constrained within $\sim 5\%$ (2.5\%).

Furthermore, considering 10 random events in ET corresponding to different glitching pulsars other than Vela (with SNR $\geq 8$), the distribution of posteriors of nuclear parameters and NS properties are shown in~\cref{fig:nuclear_10_events,fig:EOS_properties_10event}, respectively. We have tabulated the recovered  nuclear parameters resulting from this scenario in~\cref{tab:prior_posterior}. The medians and the 90\% SCIs of the reconstructed EOS and NS properties are tabulated in~\cref{tab:EOS_NS_properties}.  Though  all the nuclear parameters are well recovered, the uncertainties on the nuclear parameters are larger than what we have obtained by considering 10 events from the Vela pulsar in ET. This is expected as consideration of only Vela pulsar provides a better scenario for recovering the GW parameters due to the close distance of $\sim$300 pc. We tabulate the estimates of the reconstructed  $R_{1.4M_{\odot}}$, $f_{1.4M_{\odot}}$, and $\tau_{1.4M_{\odot}}$ in~\cref{tab:EOS_NS_properties}. With ETO10, the $f_{1.4M_{\odot}}$ can be constrained within 40 Hz.


\section{Discussions}\label{sec:conclusion}

{We revisit the inverse problem of NS asteroseismology to constrain the nuclear parameters (hence the NS EOS) within a Bayesian formalism using potential detectable GW events from f-modes in isolated NSs.
In order to estimate the detectability, in this work
we consider GW signals resulting from the excitation of f-modes in isolated pulsars, where the mode is excited to a level corresponding to the energy associated with typical pulsar glitches, as done in present f-mode searches in LIGO-Virgo data. In A+ configuration, we consider f-mode GWs from the Vela pulsar corresponding to the glitches having SNR $\geq 8$. In ET,  we considered multiple events corresponding to glitches in other pulsars, including Vela, that can be detectable due to its higher sensitivity.} For the selected events, we determine the observational uncertainties in the measurements of f-mode characteristics ($f,\tau$) and translate them to find the posterior ( hence the uncertainties ) of the nuclear parameters in a Bayesian formalism by describing the NS matter within an RMF model.

Considering a single f-mode event resulting from  Vela pulsar, in A+ (ET) we find that the nucleon effective mass ($m^*$) can be constrained within 10\%  (5\%). In both A+ and ET configurations,  among the other nuclear parameters, the compressibility ($K$) and the slope of symmetry energy ($L$) can be constrained marginally. The tighter constraint on $m^*$ is expected, as, in the considered RMF model, $m^*$ shows strong correlations with NS observables. For a single event, the uncertainty in the EOS, $M-R$ relation, $c^2_s$, and $p$ can be reduced significantly. In A+, $p$ and $c_s^2$ of the NS matter at $5n_0$ ($2n_0$) can be constrained within 25\%(35\%) and 10\% (44\%), respectively. With a detection in A+, $R_{1.4M_{\odot}}$, $f_{1.4M_{\odot}}$, and $\tau_{1.4M_{\odot}}$ can be measured within~6\%, 7\%, and 18\%, respectively. With ET, the errors in the measurements reduce significantly. For a detection in ET, $p$ and $c_s^2$ of the NS matter at $5n_0$ ($2n_0$)  can be constrained within 20\%(25\%) and 5\% (26\%), respectively.  With ET, the  $R_{1.4M_{\odot}}$, $f_{1.4M_{\odot}}$, and $\tau_{1.4M_{\odot}}$ can be measured within~4\%, 4\%, and 9\%, respectively.

Further, we consider ten multiple events of  different energies corresponding to different glitches of the Vela pulsar and investigate the impact on nuclear parameters and NS properties. We find that all nuclear parameters can be well constrained with ten events, both in A+ and ET. The $m^*$ can be constrained up to 3\%   and 2\%  in A+ and ET, respectively. We also find that the slope parameter $L$ can be constrained within $\sim$20\%. The tighter constraint on the nuclear parameters reflects on the uncertainties of EOS and NS properties compared to the resulting errors from a single event (see ~\cref{fig:EOS_properties_10event,fig:EOS_properties_1event} for comparison). With A+, $p$ and $c^2_s$ at $5n_0$ ($2n_0$),  can be constrained within ~10\% (15\%) and 3\% (15\%) respectively. However, with ET, $p$ and $c^2_s$ at $5n_0$ ($2n_0$),  can be constrained within ~5\% (6\%) and 1\% (7\%), respectively. With ten events, $f_{1.4M_{\odot}}$ can be constrained within 40 Hz and 20 Hz in A+ and ET, respectively.  Additionally, in ET, we consider 10 random f-mode GW events from different pulsars other than Vela, with SNR $\geq 8$, and investigate the inverse problem. We find that the  nuclear parameters and recovered EOS (or NS) properties have larger uncertainties than those obtained by combining ten events with different Vela glitches in the ET configuration.

{Though we consider the mode excitation with energy that of typical pulsar glitches, this methodology can be applied to other scenarios used to stimulate the NS oscillation modes: like mini collapse~\citep{Lin2011}, star quakes~\citep{Kokkotas2001}, magnetar bursts, etc.} Even without detecting electromagnetic counterparts of the associated GW events (like glitches in this work), the  GWs from the f-mode events can be localized precisely~\cite{Lopez2022} and can reveal the underlying phenomena. This work can further be improved by including the corrections due to NS rotation ~\citep{Volkel2021,Volkel2022} or magnetic field. This work assumes the NS interior consists of nucleonic matter only, which can be further improved to consider the presence of exotic matter like hyperons, quarks, and dark matter. Eq.~\ref{eqn:PE_oneevent} of~Section \ref{sec:Bayes} assumes that for an EOS model, the f-mode characteristics are single-valued as a function of $M$. However, this needs to be modified if one considers the presence of twin stars where multiple values of $f$ and $ \tau$  can occur for the same mass $M$ in a particular region. So, while considering twin stars, the Bayesian methodology needs to be modified. We will address this issue separately in our upcoming work.  Our conclusions are based on the assumption of a chosen RMF model, which can be checked with other NS EOS models, and a model comparison can be made. However, we expect that the estimated posteriors of NS properties ($M-R, R_{1.4M_{\odot}},\ f_{1.4M_{\odot}},\ \tau_{1.4M_{\odot}}$) may not change drastically as they are merely controlled by the observational uncertainty in ($f,\tau$).

\section{Acknowledgments}
The authors thank Dr. Wynn C.G. Ho for the insightful suggestions and comments on improving this work. The authors thank the anonymous referee for the useful suggestions during the review process. B.K.P. thanks Tathagata Ghosh, Divya Rana, Anuj Mishra, and Deepali Agarwal, for their discussions during this work. B.K.P. acknowledges the use of the IUCAA HPC cluster Pegasus for computational work. This work makes use of \texttt{GNU Scientific Library}, \texttt{NumPy}~\citep{Numpy} , \texttt{SciPy}~\citep{Scipy}, \texttt{StatsModel}~\citep{statsmodel},\texttt{Matplotlib}~\citep{Matplotlib}, \texttt{jupyter}~\citep{jupyter}, \texttt{dynesty}, \texttt{PyMultinest}, \texttt{bilby}, and \texttt{corner.py}~\citep{corner} software packages. 

\appendix
\section{Combining one f-mode event with the existing Astrophysical observations}\label{app:Astro_and_fmode}
 The future detection of f-modes can be combined with other astrophysical observations (or ground-based nuclear and hyper-nuclear experiments) to constrain the NS EOS. In principle, by the time of the following observation run, one would expect one or more GW events involving NS or mass-radius measurements of different pulsars to improve the understanding of the underlying physics of the dense NS matter. However, we consider the existing astrophysical observations (1) the mass and the tidal deformability measurements of the BNS event GW170817~\footnote{ Data taken from \url{https://dcc.ligo.org/LIGO-P1800115/public}}~\citep{AbbottPRL121} (2) The $M-R$ measurements of PSR J0740+6620~\citep{Miller_2021,Riley_2021} (3) The $M-R$ measurements of PSR J0030+0451
 \footnote{We consider the 3-spot $M-R$ sample data from ~\citet{miller_m_c_2019_3473466} }~\citep{Miller_2019,Riley_2019,miller_m_c_2019_3473466}   and combine them with the one futuristic event of f-mode detection from the  Vela pulsar in A+ and ET to constrain the EOS parameters. For the astrophysical observations, we follow the Bayesian methodology given in~\citet{Biswas2021} and~\cref{eqn:PE_Nevent} to estimate the posterior combining different events.

\begin{figure*}
    \centering
    \includegraphics[width=\linewidth]{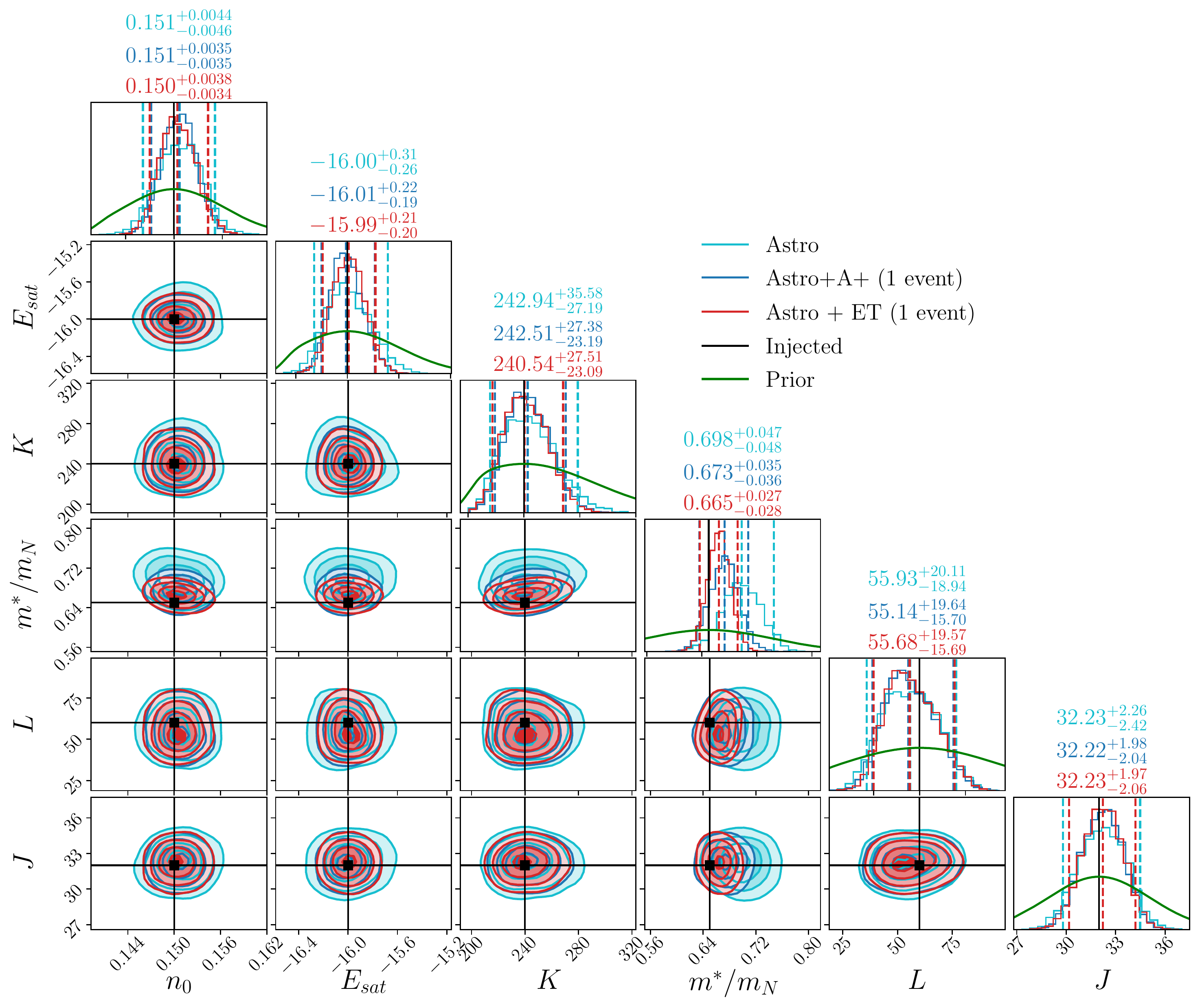}
    \caption{Similar to ~\cref{fig:nuclear_Velaglitch} but  the posteriors are resulting from considering existing astrophysical observations (cyan color), combing existing astrophysical events with 1 f-mode event in A+ (blue color) and ET (red color)   correspond to strongest vela glitch.}
    \label{fig:nuclear_1event_astro}
\end{figure*}

\begin{figure*}
\centering 

\subfigure[]{%
  \includegraphics[width=0.4\linewidth]{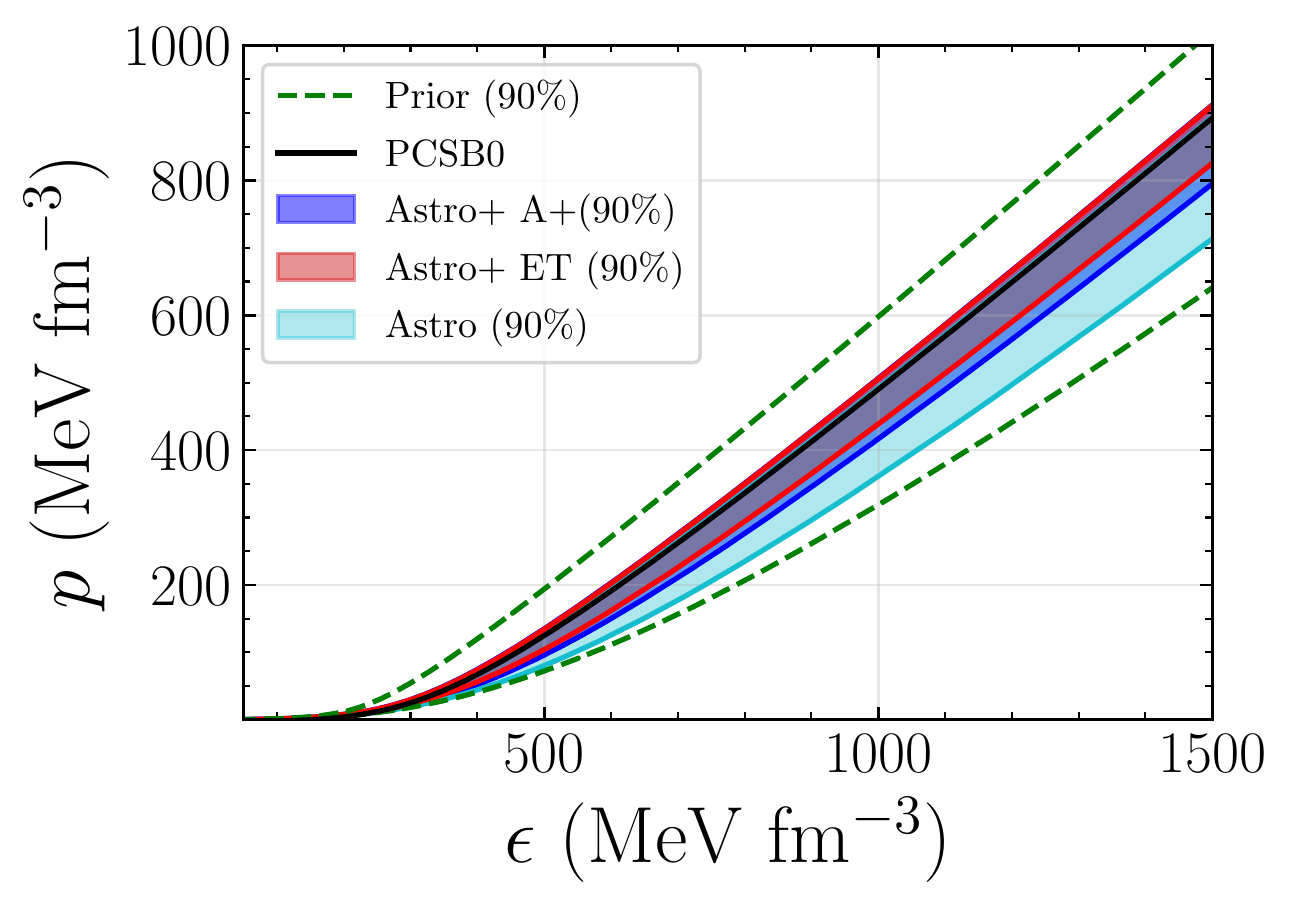}%
  \label{subfig:onewithastroevent_EOS}%
}
\subfigure[]{%
  \includegraphics[width=0.38\linewidth]{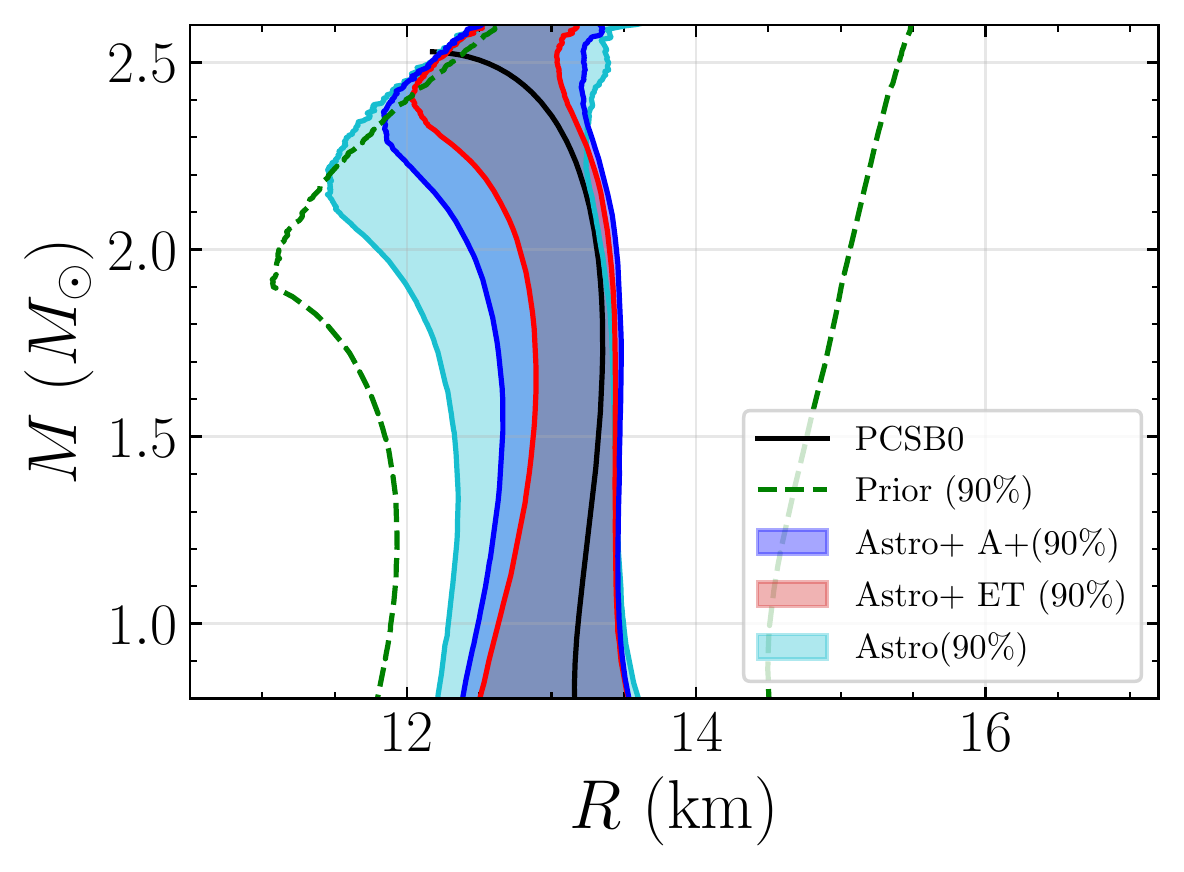}%
  \label{subfig:onewithastroevent_mr}%
}
\caption{Shows 90\% SCIs for  (a) the pressure as function of energy density and (b) the radius as a function of mass reconstructed from different posteriors of nuclear parameters presented in ~\cref{fig:nuclear_1event_astro}.}
\label{fig:EOS_properties_withastro}
\end{figure*}

 We display the posterior distributions of nuclear parameters in~\cref{fig:nuclear_1event_astro} and tabulate the median and 68\% SCI of the parameters on~\cref{tab:prior_posterior}. We display the uncertainty in the EOS and MR  measurements in~\cref{fig:EOS_properties_withastro}  and tabulate the medians and 90\% SCI of few EOS and NS properties in~\cref{tab:EOS_NS_properties}. From~\cref{fig:nuclear_1event_astro}, one can conclude that after adding the f-mode events, the uncertainty in the posterior of $m^*$ improved significantly compared to the posterior of $m^*$ obtained with the existing astrophysical observations. Whereas the resulting uncertainties of the nuclear parameters other than $m^*$, obtained after adding the f-mode detected event, are marginally improved compared to those obtained by combining the existing astrophysical observations only. However, the tighter constraint on $m^*$ after adding an f-mode event improves the 90\% SCI of NS EOS or the $M-R$ uncertainties (see~\cref{fig:EOS_properties_withastro}), indicating a better constrain. The uncertainties in the measurements of EOS or $M-R$ improved, considering the event being detected by ET as compared to A+.

\bibliography{Pradhan}{}
\bibliographystyle{aasjournal}

\end{document}